\documentclass[a4paper,11pt]{article}
\pdfoutput=1 

\usepackage{jcappub} 

\usepackage[T1]{fontenc} 
\usepackage{amsmath}
\usepackage{amssymb}
\usepackage{gensymb}
\usepackage[referable]{threeparttablex}
\usepackage{booktabs, longtable}

\usepackage{grffile}  

\newcommand{\be}{\begin{equation}}
\newcommand{\ee}{\end{equation}}
\newcommand{\diff}{\ensuremath{\mathrm{d}}}

\usepackage{xcolor}

\definecolor{darkBlue}{rgb}{0, 0, 0.8}

\definecolor{dark-red}{rgb}{0.0,0.0,0.0}
\definecolor{green}{rgb}{0.0,0.0,0.0}
\definecolor{DarkBlue}{rgb}{0.0,0.0,0.0}

\renewcommand{\vec}[1]{\boldsymbol{#1}}

\title{Explaining the UHECR spectrum, composition and large-scale anisotropies with radio galaxies}

\author[a,b,1]{B.~Eichmann,\note{Corresponding author.}}
\author[a]{M.~Kachelrie\ss}
\author[a]{and F.~Oikonomou}

\affiliation[a]{Institutt for fysikk, 
  NTNU Trondheim, Norway}

\affiliation[b]{Ruhr Astroparticle and Plasma Physics Center (RAPP Center), Ruhr-Universit\"at Bochum, Institut f\"ur Theoretische Physik IV, 44780 Bochum, Germany}


\abstract{
Radio galaxies are promising candidates as the sources of ultrahigh energy
cosmic rays (UHECRs). In this work, we examine if the stringent constraints
imposed by the dipole and quadropole anisotropies as well as the UHECR
spectrum and composition allow that radio galaxies are the dominant
extragalactic cosmic ray sources. In order to calculate the UHECR flux
emitted by individual
radio galaxies, we constrain their properties using information from the
radio-CR correlation and a dynamical evolution model. In
addition to the UHECR flux from individual, local sources, we 
include the diffuse flux emitted by the bulk of non-local radio galaxies based
on their radio luminosity distribution. Analyzing the source parameters within
a range around their expected properties, we finally determine the
configurations of local sources describing well the UHECR spectrum, composition
and large-scale anisotropies. We obtain a good description of all data
even in the case that we include only a small number of local sources. In
particular, we find that scenarios where few sources like Fornax~A and Virgo~A
dominate the flux above the ankle, while low-luminosity radio galaxies
contribute an isotropic background dominating below the ankle,
provide a good fit to the data.
}

\begin{document}
\maketitle
\flushbottom

\section{Introduction}
\label{sec:intro}

The origin of cosmic rays with energy exceeding $10^{18}$~eV, termed here
ultra-high energy cosmic rays (UHECRs), is a long-standing mystery. In the
last 15~years, considerable progress has been made in characterising the
diffuse
spectrum~\cite{TelescopeArray:2012qqu,PierreAuger:2020qqz,PierreAuger:2020kuy},
and its elemental composition~\cite{Yushkov:2020nhr,Hanlon:2020wbz},
as a result of the observations made with the Pierre
Auger Observatory and the Telescope Array experiment. A recent breakthrough
in the study of UHECR anisotropies has been the discovery of a dipolar
anisotropy with an amplitude $\delta\simeq 6.5\%$ in the arrival directions
of UHECRs exceeding $8\times10^{18}$~eV~\cite{Aab:2017tyv}. While the best-fit
value of this amplitude  increases as $\delta\propto E^{0.8}$ with
energy~\cite{PierreAuger:2018zqu}, neither the dipole moments at higher
energies nor the quadrupole moments have at present more than $3\,\sigma$
significance~\cite{TelescopeArray:2021ygq}. In addition, several hints for
correlation of the UHECR arrival directions with extragalactic $\gamma$-ray
emitting sources at smaller
angular scales exist~\cite{Aab:2018chp}. Despite this progress,
many questions remain open, see e.g.\ Refs.~\cite{Mollerach:2017idb,AlvesBatista:2019tlv,Kachelriess:2019oqu,Kachelriess:2022phl} for recent reviews. The most pressing one is
certainly the determination of the sources of these extremely energetic
particles.

Among the few  potential source candidates, which appear to satisfy the
necessary conditions for accelerating cosmic rays to ultra-high
energies~\cite{1984ARA&A..22..425H}, AGN with collimated jets, also referred
to as ``radio galaxies'', are one of the most promising. These sources exhibit
sufficiently high power as to plausibly be able to accelerate
UHECRs~\cite{1976Natur.262..649L}, large size and longevity. Furthermore,
their emissivity is sufficient to explain the inferred UHECR emissivity
which is approximately $6 \times 10^{44}\,{\rm erg}/{\rm Mpc^{3}/yr}$
for energies $E>5\times 10^{18}$\,eV~\cite{PierreAuger:2020qqz}, i.e.\ above
the ankle
(see Ref.~\cite{Kachelriess:2022phl} for a recent review of these constraints). 
With an exceptionally small distance of about 4\,Mpc, the radio galaxy
Centaurus~A (NGC~5128) is by far the closest AGN with respect to
Earth. In addition, it is one of the brightest jetted AGN in terms of radio
flux and has long been suggested as a likely UHECR source~\cite{1996APh.....5..279R,Kachelriess:2008qx,2009MNRAS.393.1041H,2009ApJ...706.1517H,2009A&A...506L..41R,Kachelriess:2010zk,2010ApJ...720L.155G}. 
In recent years, there have been multiple observational hints~\cite{PierreAuger:2014yba,Caccianiga:2019Sr}, 
and model results~\cite{Eichmann+2018, Matthews+2018, MollerachRoulet2021arXiv}
indicating that Centaurus~A may be responsible for at least some fraction of the
observed UHECR flux. Apart from Centaurus~A,  also Fornax~A (NGC~1316) and
Virgo~A (NGC~4486) stand out in terms of their extraordinarily high radio flux
in the local ($<21$\,Mpc) Universe.  Adding these sources yields also some
benefit in explaining the observed arrival
directions~\cite{Matthews+2018,deOliveira:2021ckh}. 
In addition, several studies have demonstrated that the UHECR spectrum and composition 
are consistent with UHECRs originating in radio 
galaxies~\cite{Kimura:2017ubz,Fang:2017zjf,Rodrigues:2020pli}. However, these studies
do neither account for the local distribution of radio galaxies nor for the data on
the UHECR arrival direction distribution. Therefore, no complete
explanation of all UHECR data, i.e.\ the spectrum, elemental composition,
and arrival directions, accounting for the dominance of these local sources
has been suggested so far.
Several authors have investigated whether the observed large-scale anisotropy
of UHECR arrival directions is consistent with arising from the non-uniform
matter distribution in the local Universe, taking into account the observed
UHECR composition, and the deflections of UHECRs in the Galactic and
extragalactic magnetic fields. In the recent analyses of
Refs.~\cite{Ding:2021emg,Allard+2021arXiv}, it was shown that such a scenario
is consistent with the amplitude and energy dependence of the observed dipole
anisotropy. The direction of the dipole is more difficult to reproduce, but
also strongly affected by the uncertainties of the Galactic magnetic field.
In particular, Allard et al.~\cite{Allard+2021arXiv} found that the majority of
their simulated scenarios exhibit larger quadrapolar anisotropies than observed,
challenging the scenario that UHECRs are accelerated in a large number of
UHECRs sources which follow the large-scale structure.

A major obstacle for the identification of the UHECR sources are the
deflection of these particles by the not well known Galactic and
extragalactic magnetic fields (hereafter GMF and EGMF, respectively). In
addition to hiding the sky position of the sources, these deflections
can in case of a strong EGMF also cause significant time delays that exceed
the lifetime of the potential sources. As the potential UHECR sources are
commonly selected based on their electromagnetic properties, large time
delays would undermine such a correlation. Moreover, many AGN show
a varied and complex history~\cite{Konar+2019,Maccagni+2020,Croston:2009zc},
so that the actual source properties at a given time can significantly
deviate from the time-integrated ones. In that sense, the observed
source luminosity is not necessarily an
accurate tracer of the total UHECR power of the source, since only
fast variability ($\lesssim 1\,\text{Myr}$) is smoothed
out~\cite{MatthewsTaylor2021}. In case of variability on longer timescales,
it has been shown for typical source properties that the luminosity in the
radio band is the most reliable proxy for the UHECR luminosity
\cite{MatthewsTaylor2021}. Furthermore,
these arguments indicate that it is not reasonable to assume that the UHECR
sources have identical properties.

In this work, we examine if the stringent constraints imposed by the data on
the dipole and quadropole anisotropies as well as the UHECR spectrum and
composition allow that radio galaxies are the dominant extragalactic cosmic
ray sources. Our aim is to go beyond the ``average source'' approach and to
model the properties of local sources individually.  In order to calculate
the UHECR flux emitted by individual radio galaxies, we assume the validity
of the radio-CR correlation to estimate properties like their cosmic ray power
and the maximal rigidities of the accelerated particles. 
Moreover, we account for the finite lifetime of these sources, e.g., by the
inclusion of a dynamical evolution model of high-luminosity radio sources. 
Since the uncertainties of these parameters are rather large, we do not fix
them to unique values but allow them to float within a prescribed range to
explain the observational data.
For the calculation of the UHECR energy spectrum, composition and
aniostropies at Earth, we propagate UHECRs
using a combination of a semi-ballistic and diffusive description:
We describe the energy losses as a continuous change in rigidity during
one-dimensional propagation, while deflections in the turbulent EGMF are
modelled by Fisher distributions
following the approach of Refs.~\cite{MollerachRoulet2019, Harari+2021}.
Then we use the Janson-Farrar (hereafter JF12) model~\cite{JanssonFarrar2012}
to describe the deflections by the Galactic magnetic field. 
Finally, we add a diffuse flux component emitted by the bulk of radio
galaxies based on their radio luminosity distribution~\cite{Eichmann2019}.
In general, we obtain a good description of the UHECR spectrum, its composition
and the dipole and quadrupole anisotropies, even using only
a small sample of the most powerful local sources combined with an isotropic
contribution from low-luminosity radio galaxies.

This paper is structured as follows: In Sec.~\ref{sec:Ind}, we first describe 
the connection between the radio-CR correlation and the initial rigidity
spectra emitted by individual, local radio galaxies. Then we present
our modelling approach of UHECR energy losses during propagation and how these
modify the rigidity spectra, before we discuss how we treat deflections
in magnetic fields
and present finally the flux of the bulk of radio galaxies.
In Sec.~\ref{sec:FitResults}, we first discuss how we convert
the rigidity spectra to energy spectra. Next we review  our
fit parameters and procedure, and present then the fit results.
Finally,  we discuss in Sec.~\ref{sec:disc} the results and the
underlying assumptions, before we conclude in Sec.~\ref{sec:concl}.

\section{UHECRs: From the source to the signal}
\label{sec:Ind}

The emission of UHECRs is expected to vary drastically even within a single
source class, with  differences caused, e.g.,  by intrinsic differences of
their power supply and their different evolutionary stages.
Connecting the observed UHECR intensity with properties of their sources
requires therefore to abandon the idea of representative ``average sources'',
at least for our local environment. We first describe how we determine
the properties that dictate the UHECR emission of individual radio
galaxies, before we discuss our approach for the propagation and
the deflections of UHECRs.

\subsection{Radio-CR correlation}
\label{RadioCR}

The power $Q_{\rm cr}$ emitted in form of CRs by an individual jetted AGN
can be related to its jet power $Q_{\rm jet}$ as
\be
Q_{\rm cr}=\frac{g_{\rm m}}{1+k}\,Q_{\rm jet} ,
\label{Qcr}
\ee
where $g_{\rm m}$ denotes the fraction of jet energy in relativistic particles
and $k$ the ratio of leptonic to hadronic energy density~\cite{Eichmann2019}.
Generically, one expects $g_{\rm m}<1$ and in the special case of equipartition
between the  energy in leptons, hadrons and the magnetic field one finds
$g_{\rm m}\simeq 4/7$ \cite{1970ranp.book.....P}.
The jet power $Q_{\rm jet}$ can be in turn related to the observed radio
emission introducing the so-called radio-jet power correlation,
e.g.~\cite{Willott+1999,GodfreyShabala2013,Ineson+2017},
\be
 Q_{\rm jet} = Q_0\,\left( \frac{L_{151}}{L_{p}} \right)^{\beta_L},
\label{Qjet}
\ee
a relation
that provides an estimate of the jet power based on the radio luminosity
$L_{151}$ at 151\,MHz for the majority of radio-loud AGN.
Note that $Q_{\rm jet}$
of individual sources can deviate up to an order of magnitude from this
correlation. Moreover, there is a lack of reliable empirical methods to
measure the kinetic jet power---especially in the case of low-luminosity
galaxies, such as FR-I galaxies, since their energy budget seems to be
dominated by non-radiating particles~\cite{Croston+2008, Birzan+2008}.
As a result, the X-ray cavity method~\cite{Cavagnolo+2010}, that is applied
to estimate the FR-I jet power, bears comparably large uncertainties.
In contrast, the internal pressure in FR-II sources can be gauged
rather accurately, by estimating the lobe volume as well as using the radio
and X-ray flux from synchrotron and inverse-Compton emission~\cite{Ineson+2017}.
Despite these differences, Godfrey and Shabala~\cite{GodfreyShabala2013}
found approximate agreement between the radio--jet power correlation of
FR-I and FR-II sources. Therefore, we will apply the same
correlation~(\ref{Qjet}) for both low- and high-luminosity radio galaxies,
employing the parameters from the most recent analysis by Ineson et
al.~\cite{Ineson+2017} with  $\beta_L=0.89$ as slope and
$Q_0=5\times 10^{46}\,\mathrm{erg/s}$ as normalization at the pivot
luminosity $L_p=10^{28}\,\mathrm{W/Hz/sr}$. Within a few tens of Mpc, Centaurus~A ($L_{151}=1.4\times10^{23}\,\mathrm{W/Hz/sr}$), Virgo~A ($L_{151}=1.9\times10^{24}\,\mathrm{W/Hz/sr}$) and Fornax~A ($L_{151}=3.0\times10^{24}\,\mathrm{W/Hz/sr}$) feature the highest radio luminosity, leading to jet powers of a few $\times 10^{43}\mathrm{erg}\,\mathrm{s}^{-1}$, that is about an order of magnitude smaller than their expected cavity power \cite{Matthews+2018} and closer to the expected jet powers that one obtains from enthalpy calculations of their thermal pressure, e.g.~Ref.~\cite{Russell+2013}.

On large scales, i.e.\ on distances $\gtrsim 1\,\text{pc}$, it has been shown in previous works, e.g.~Ref.~\cite{Eichmann+2018,MatthewsTaylor2021}, that the escape time $\tau_{\rm esc} \simeq l/(\beta_{\rm sh}c)$ dominates over the energy loss timescales. Here, we suppose that the escape is set by the advection according to the shock or shear velocity $\beta_{\rm sh}c$ and the characteristic length scale $l$ of the jet, but dependent on the given magnetic field structure UHECRs could also diffuse out of these jets more quickly than they are advected away.   
For the common assumption of Bohm diffusion, the acceleration timescale is
given by $\tau_{\rm acc}=f_{\rm diff}\,r_{\rm L}/(c \beta_{\rm sh}^2)$
for cosmic rays  with rigidity $R=cp/Ze$ and Larmor radius $r_{\rm L}=R/B$.
Here, the parameter $1\lesssim f_{\rm diff}\lesssim 8$ encapsulates the
variations due to different shock and magnetic field
geometries~\cite{Drury:1983zz}.
In steady state, the equality of both time scales yields the maximal rigidity
\be
\hat{R}\equiv \frac{E_{\rm max} }{ Z\,e} =
\frac{\beta_{\rm sh} }{ f_{\rm diff}}\,B\, l =
g_{\rm acc} \sqrt{\frac{(1-g_{\rm m})Q_{\rm jet}}{c}}\,,
\label{Rmax}
\ee
where we introduced the magnetic field power of the jet,
$Q_B=c\,\beta_{\rm jet}\,\pi l^2\,B^2/8\pi = Q_{\rm jet}(1-g_{\rm m})$ and
the acceleration efficiency parameter
\be
g_{\rm acc}=\sqrt{\frac{8\,\beta_{\rm sh}^2}{f_{\rm diff}^2\,\beta_{\rm jet}}} .
\ee
For typical values of the  shock and jet velocities
in extended jets of radio galaxies, this  parameter lies in the range
$0.01 \lesssim g_{\rm acc} \lesssim 1$.

Jets of radio galaxies are suitable sites for first-order Fermi
acceleration at  internal or external
shocks~\cite{1981heaa.book.....L, Biermann:1987ep, Rachen:1992pg, Mannheim:1993jg, Matthews+2019}, or by shear at the jet boundary~\cite{1998A&A...335..134O,Webb+2020}.  While first-order Fermi acceleration predicts in the test particle
picture power-law spectra with slope $\alpha=2$  for strong, non-relativistic
shocks, a variety of effects like the back-reaction of CRs on the shock,
an energy-dependent escape and relativistic effects  modify the spectra.
Moreover, additional acceleration mechanisms like reconnection may change
the energy spectrum of accelerated particles. We use therefore, motivated
by simplicity, a power law for the initial differential injection rate as
function of rigidity of CR nuclei of type $i$,
\be
S_{0,i}(R_0)\equiv \frac{\diff N_{\rm cr,i}}{\diff R_0\,\diff t} = s_0\,\left(\frac{R_0}{\check{R}} \right)^{-\alpha}\,\exp[-R_0/\hat{R}] ,
\ee
but keep the slope $\alpha$ as a free fit parameter within the range
$1.5 < \alpha < 2.5$.
These rigidity distributions can be normalized by the total CR power, 
\be
Q_{\rm cr}=\sum_i Q_{\rm cr,i}\,,\quad\text{and}\quad 
Q_{\rm cr,i}=Z_i\int_{\check{R}}^{\hat{R}}\diff R_0\,\,R_0\,\frac{\diff N_{\rm cr,i}}{\diff R_0\,\diff t}\,.
\ee
Using for simplicity a Heaviside step function instead of an exponential cutoff at $\hat{R}$ for the normalisation
condition gives
\be
s_0 = \frac{Q_{\rm cr,i}}{Z_i}\, \frac{\check{R}^{-\alpha}\,(2-\alpha)}{(\hat{R}^{2-\alpha}-\check{R}^{2-\alpha})}\,.
\ee
According to the thermal-leakage models, e.g.~\cite{Bell1978b, EllisonMoebius1987, Caprioli_Spitkovsky_2014, Caprioli+2017}, it is expected that CR species thermalize at a temperature proportional to their mass~\cite{Caprioli+2017} such that $p_{\rm inj}\sim 4 \,A_0m_{\rm p}\,\beta_{\rm sh}c$, and 
\be
 \check{R}=\frac{cp_{\rm inj}}{eZ_0} \sim 1\,\text{GV} \,,
\ee
where we used $A_0\simeq 2\,Z_0$ for the numerical value. Therefore, the minimal rigidity is expected not to depend on the initial particle type of the CR. Note that the dependence on $\check{R}$ practically vanishes
in the case of $\alpha<2$ because of $\hat{R}\gg\check{R}$.

\subsection{UHECRs from individual radio galaxies}

\begin{figure}[htbp]
\centering
\includegraphics[width=.420\linewidth]{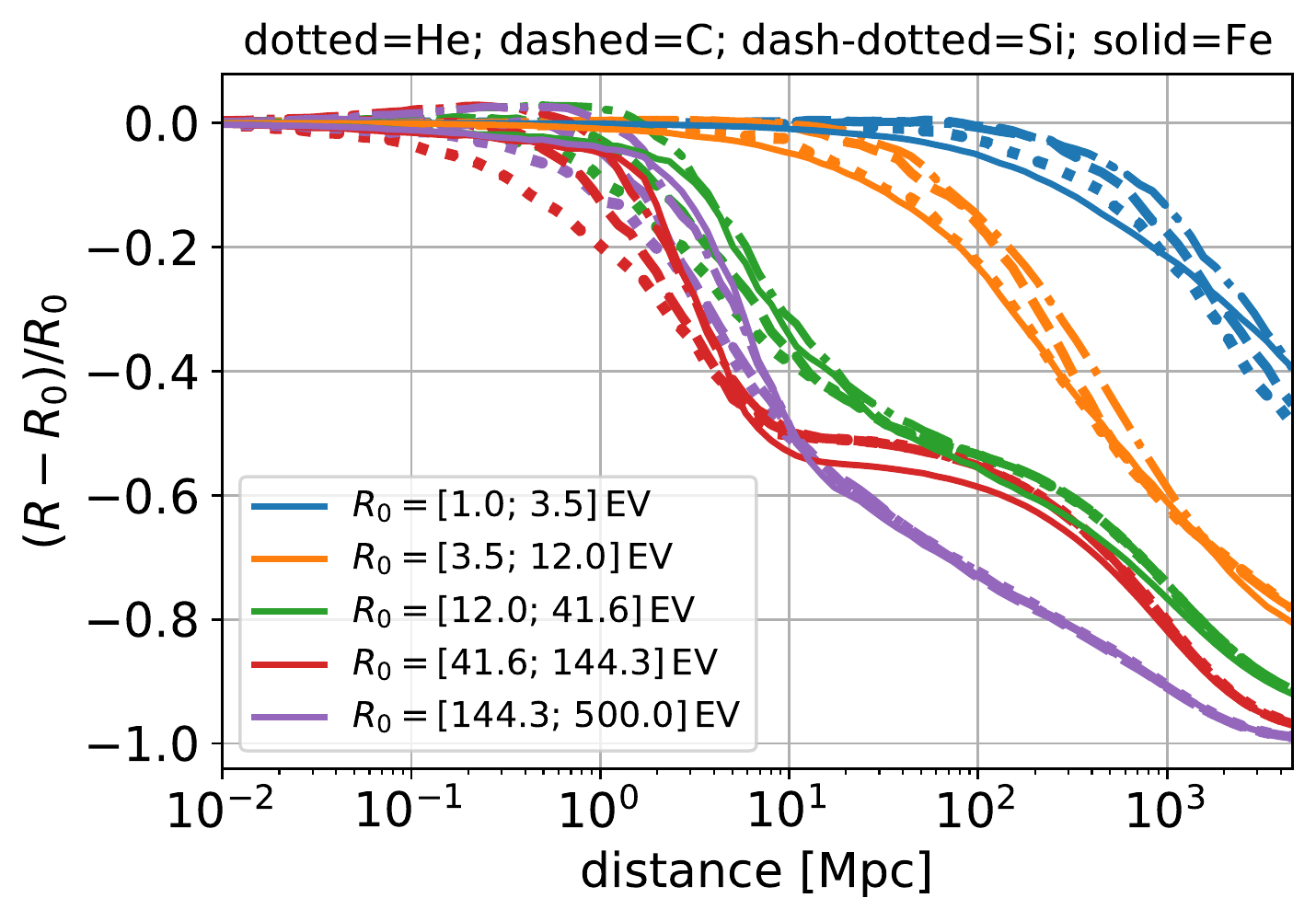}
\includegraphics[width=.570\linewidth]{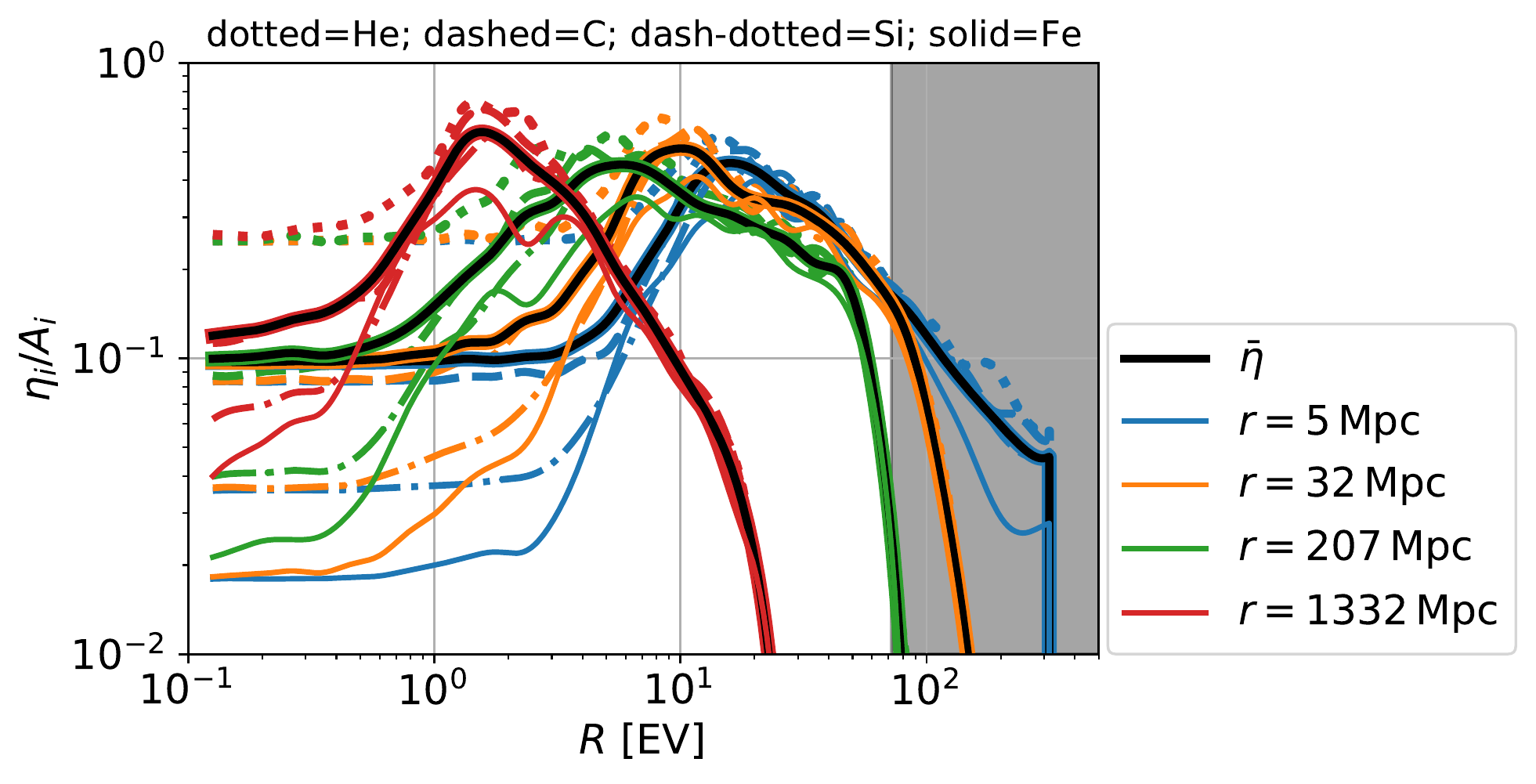} 
\caption{The impact of interactions with the CMB and the EBL on the CR properties:
\emph{Left}: The change of rigidity of different primary particle types as a function of the propagation distance. 
\emph{Right}: The modification factor $\eta_i$ of different nucleus types (coloured lines) as well as their average (black solid line) for different source distances as a function of the rigidity for a source spectral index $\alpha\simeq 2$ and $\hat{R}=72\,\text{EV}$ (as indicated by the black shaded area).
}
\label{fig:interactionEffects}
\end{figure}

The propagation of UHECRs from the sources to Earth is on the one hand
affected by  deflections in magnetic fields and on the
other hand by interactions with photons from the cosmic microwave background
(CMB) and the extragalactic background light (EBL). While the former only
depend on the particle's rigidity, the latter also depends on the mass number
$A$ of the CR nucleus. However, the change of rigidity $(R-R_0)/R_0$ of the
primary CR depends rather weakly on the nucleus type, as shown in the left
panel of Fig.~\ref{fig:interactionEffects}. We will therefore use rigidity
as the evolution variable measuring the energy losses of UHECRs
during propagation.

The differential number density of CRs  of type $i$ at the present time $t_0$ can be determined as
\be
n_{i}(R,r) =\int_0^{t_0} \diff t\,\, S_{0,i}[R_0(R,t),t]\,P_i(R,t,r)\,\frac{\diff R_0}{\diff R}(R,t) ,
\label{eq:diffNumDensCR_ABG}
\ee
where $P_i$ is the ``propagator'' corresponding to the Green function
of the relativistic diffusion equation including continuous energy
losses~\cite{AloisioBerezinskyGazizov2009}.
Numerical simulations of UHECR sources with finite lifetime $t_{\rm act}$  at
small enough distances $r$, such that energy losses can be neglected,
yield~\cite{MollerachRoulet2019, Harari+2021}
\be
 n_{i}(R,r,t_{\rm act}) = \frac{S_{0,i}(R)}{4\pi\,c\,r^2}\,\xi(R,r,t_{\rm act}) 
\label{eq:diffNumDensCR0}
\ee
with the enhancement factor~\cite{Harari+2021}
\be
\xi(R,r,t_{\rm act}) \simeq
\frac{1}{C(R,r)} 
\exp\left[-\left( \frac{r^2}{0.6l_{\rm D}(ct_{\rm act}+r)} \right)^{0.8}  \right] .
\label{eq:enhancement}
\ee
Here, the prefactor
\be
C(R,r) = \frac{l_{\rm D}}{3r}\,\left[ 1-\exp\left(-3r/l_{\rm D}-3.5(r/l_{\rm D})^2 \right) \right] 
\ee
is determined by the ratio of the source distance $r$ and the diffusion
length
\be
l_D \simeq l_{\rm c}\left[ 4\left( \frac{R}{B\,l_{\rm c}} \right)^2 + 0.9\left( \frac{R}{B\,l_{\rm c}} \right) + 0.23\left( \frac{R}{B\,l_{\rm c}} \right)^{1/3}\right]\,,
\ee
where we use the fit obtained for an isotropic Gaussian random field with a coherence length $l_{\rm c}$ and a Kolmogorov spectrum from
Refs.~\cite{Parizot:2004wh, Harari+2014}. 
Note that $C(R,r)\simeq 1$ in the case of rectilinear propagation, i.e.\ for
$r\ll l_{\rm D}$, as well as $\xi(R,r,t_{\rm act})\simeq 1$ if also short
lifetimes, $t_{\rm act}\ll 2l_{\rm D}/c$, are considered.

In this work, we define the source lifetime $t_{\rm act}$ with respect to the
observation at Earth. Hence, the maximal distance traveled by the observed
UHECRs equals $ct_{\rm act}+r$. Equation~(\ref{eq:diffNumDensCR0}) was derived in
the limit that energy losses can be neglected. Thus it does not account
for the production of secondary nuclei or the change of rigidity in general.
For sources with a distance above tens to hundreds of Mpc dependent on the
initial CR rigidity, the left panel of
Fig.~\ref{fig:interactionEffects} indicates that these approximations no
longer hold. Therefore, we introduce a modification parameter $\eta_i$ that
is defined as the ratio of the rigidity spectrum, $S_{i}(R,r)$ including
interactions with the CMB and the EBL and the initial source spectrum
(i.e. in the absence of interactions),
$S_{0,i}(R)$, so that
\be
\eta_i(R,r) \equiv \frac{S_{i}(R,r)}{S_{0,i}(R)} .
\ee
Using the open-source package CRPropa\,3~\cite{CRPropa3_2016, CRPropa3.1_2017},
we determine $\eta_i(R,r)$ by 1D simulations of the propagation of He, C, Si
and Fe nuclei for a grid of values of the  spectral index $\alpha$ and the
maximal rigidity $\hat{R}$. 

As shown in the right panel of Fig.~\ref{fig:interactionEffects}, $\eta_i$ in
principle depends on the initial nucleus type $i$.  However, at high rigidities
and/or large source distances it converges towards $A_i$, corresponding to
the total photo-disintegration of the initial nucleus of mass number $A_i$,
modified by the exponential cutoff at the maximal rigidity. Although there is
still some weak dependence on the nucleus type at rigidities
$\lesssim 10\,\text{EV}$, we will subsequently use the approximation that
\be
\eta_{i}(R,r)\simeq A_i\,\bar{\eta}(R,r) \equiv A_i\,\langle \eta_{i}(R,r) / A_i \rangle_i .
\label{eq:eta_bar}
\ee
Furthermore, we adopt for the calculation of $\eta_i$ an equipartition of He,
C, Si and Fe nuclei, which corresponds to a mean mass number $\bar{A}=25$ at
a given rigidity. 
Note that the chosen spectral index $\alpha$ has a comparably small impact
on the spread of $\eta_{i}(R,r) / A_i$ as function of the CR type $i$,
and in general this spread becomes smaller with decreasing $\alpha$.
As an example, we note that for $R=1\,\text{EV}$ and $r=100\,\text{Mpc}$
the difference of the relative spread
$[\eta_{He}(R,r)/4-\eta_{Fe}(R,r)/56]/(\eta_{He}(R,r)/4)$ between $\alpha=2$
and $\alpha=1.5$ is about $0.1$, hence, much smaller than the relative
difference between $\bar{\eta}$ and $\eta_{i}(R,r) / A_i$ in most cases.
Since the general spectral behavior of $\eta_{i}(R,r) / A_i$ is quite similar
for all nuclei types, the relative difference of the actual modification
factor $\eta_{i}(R,r) / A_i$ between two sources that feature similar elemental
abundances are typically much smaller than a factor of five. 

\begin{figure}[htbp]
\centering
\includegraphics[width=.495\linewidth]{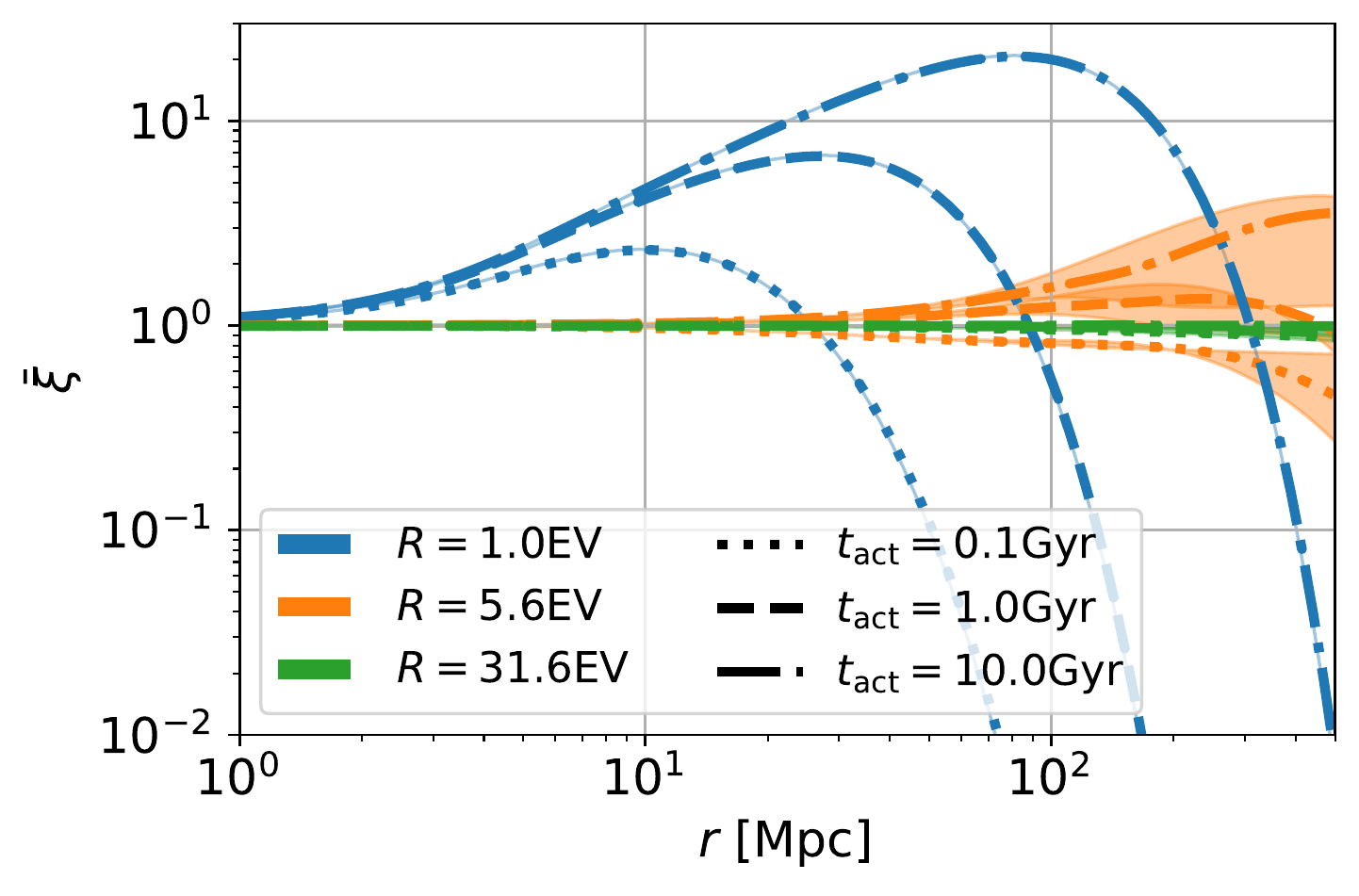}
\includegraphics[width=.495\linewidth]{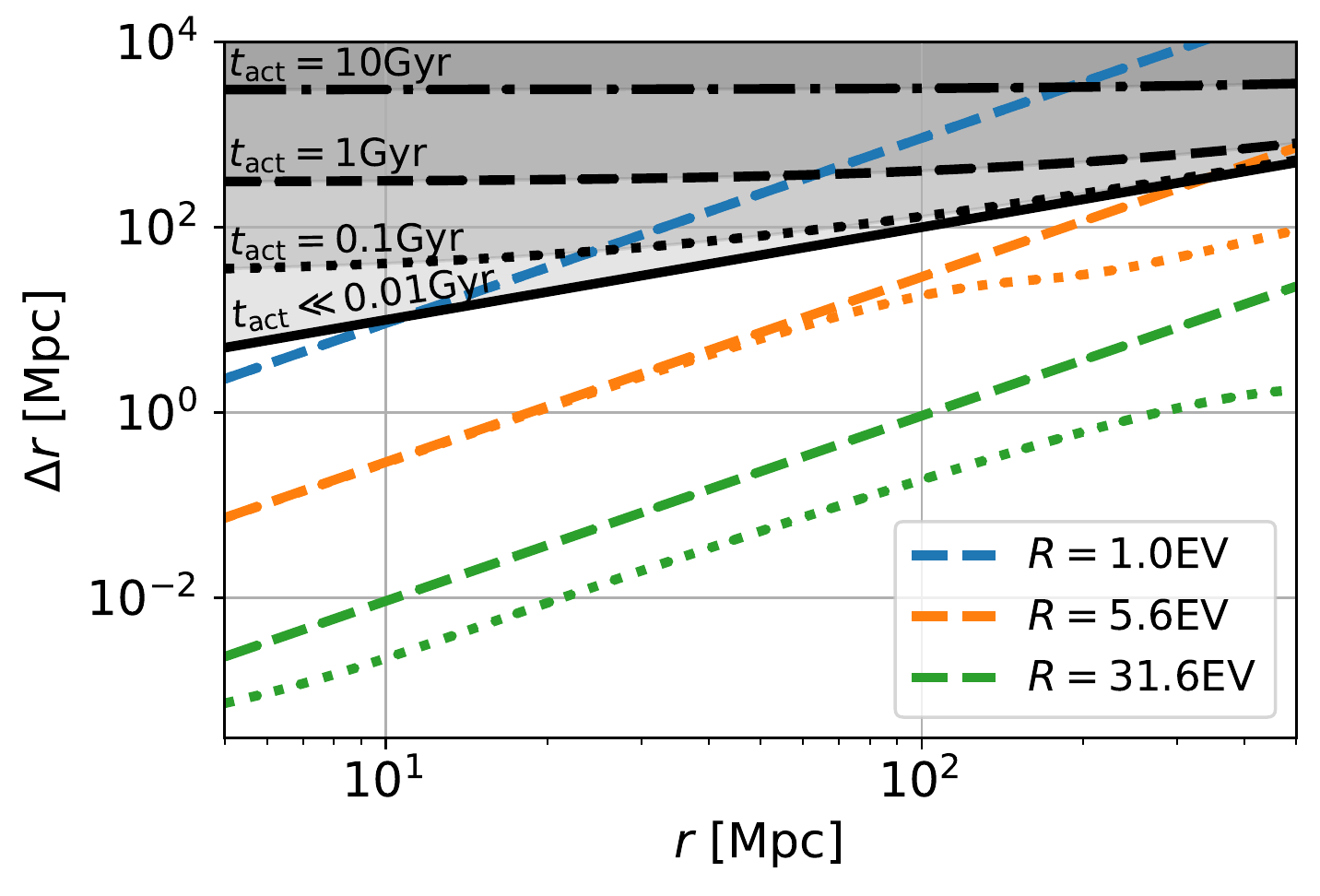} 
\caption{\emph{Left}: The mean enhancement factor \ref{eq:mean_enh} as a function of the source distance for three different source lifetimes and rigidities. The edges of the coloured bands indicate the corresponding enhancement factor \ref{eq:enhancement} for $R$ and $R_0(R,r)$, respectively.
\emph{Right}: The additional mean propagation length $\Delta r$ due to isotropic diffusion by the EGMF, where we use $B_{\rm rms}=1\,\text{nG}$ and $\lambda=1\,\text{Mpc}$. The dotted colored lines correspond to the resulting propagation delay for a rigidity $R_0(R,r)$ at the sources. The black lines indicate the maximal propagation length due to a finite source lifetime.
}
\label{fig:propDelay_enh}
\end{figure}

Including the modification factor $\eta_i$, the CR density is determined as
\be
n_{i}(R,r,t_{\rm act}) = \frac{S_{0,i}(R)}{4\pi\,c\,r^2}\,\xi(R,r,t_{\rm act})\,A_i\,\bar{\eta}(R,r).
\label{eq:diffNumDensCR1}
\ee
Note that this approximation does not include the increased length of the
propagation distance due to the deflections by the EGMF or the change of
the diffusion length $l_D$ with distance due to the change of rigidity.
The effect of the latter is roughly taken into account by using
the average enhancement factor
\be
\bar{\xi}(R, r,t_{\rm act}) \equiv \langle \xi(R_0(R,x), r,t_{\rm act}) \rangle_x = \frac{1}{r}\int_0^r \diff x\,\,\xi(R_0(R,x), r,t_{\rm act}) .
\label{eq:mean_enh}
\ee
A comparison with the enhancement factor~(\ref{eq:enhancement}) for the
limiting rigidities of $R_0$ and $R$, respectively, shows that
$\bar{\xi}\simeq \xi(R_0(R,r)) \simeq \xi(R)$ is valid for a very large range of
parameters, i.e.\ where $l_{\rm D}\, {}^{\gg}_\ll\, r$ or
$t_{\rm act}c \lesssim r$, so that averaging has only a minor impact on the
results (cf.\ the left panel of Fig.~\ref{fig:propDelay_enh}). In the
case of small rigidities (blue lines) the change of rigidity is negligible,
hence $R\simeq R_0$, and for large rigidities (green lines) $l_{\rm D} \gg r$
so that the argument of the exponential term in Eq.~(\ref{eq:enhancement})
vanishes. For CR rigidities $\gtrsim 6\,\text{EV}$, EGMF strengths of
$B_{\rm rms}\lesssim 1\,\text{nG}$ and a coherence length of $1\,\text{Mpc}$,
the right panel of Fig.~\ref{fig:propDelay_enh} illustrates that UHECRs from
sources up to a few hundreds of Mpc have a diffusive propagation delay which is
negligible compared to their source distance. For a significantly stronger
EGMF and/or lower CR rigidities, there is only a small range of source
distances where the average propagation delay is significantly larger than
the source distance but smaller than the maximal propagation distance,
i.e.\ $r<\Delta r<t_{\rm act}c+r$. Hence, in most cases the majority of UHECRs
have on average a propagation length quite close to their source distance,
so that it is reasonable to neglect the additional propagation length at
zeroth order. Note that this holds even stronger in the case of quasi-ballistic propagation, where $\Delta r \propto r$, which e.g.\ leads to a flattening of $\Delta r(R=5.6\,\text{EV})$ at about 30\,Mpc.

Finally, with $A_i\simeq 2\,Z_i$ the total CR density becomes
\be
n(R,r,t_{\rm act}) = \sum_i n_{i}(R,r,t_{\rm act}) = n_0
\,\left(\frac{R}{\check{R}} \right)^{-\alpha}\,\exp\left(-\frac{R}{\hat{R}}\right) \,\bar{\xi}(R,r,t_{\rm act})\,\bar{\eta}(R,r) ,
\label{eq:totDiffNumDensCR}
\ee
where
\be
n_0 = \frac{Q_{\rm cr}}{2\pi c\,r^2}\, \frac{\check{R}^{-\alpha}\,(2-\alpha)}{(\hat{R}^{2-\alpha}-\check{R}^{2-\alpha})}
\ee
does \emph{not\/} depend on the initial nucleus type. Thus, we are able to
compare our model predictions to observational data (see
Sect.~\ref{sec:FitResults}) without the need to specify the initial
abundances of elements at the sources, which implies
a substantial reduction
of the dimensionality of the parameter space we have to consider\footnote{Note,
  that in principle each of the $N_{\rm src}$ sources (or at least source
  classes) can contribute an individual set of $N_{\rm i}$ elemental abundances
  leading to $(N_{\rm i}-1)\times N_{\rm src}$ free parameters.}.

\subsubsection{Local radio galaxies}

For a given total UHECR density~(\ref{eq:totDiffNumDensCR}) at a distance $r_s$
from an individual, local source $s$, the corresponding total UHECR intensity
as function of rigidity is given by
\be \label{diff}
J_s(R,r_s,t_{\rm act})\equiv \frac{\diff N_{\rm cr}}{\diff R\,\diff A\,\diff t\,\diff\Omega} = \frac{c}{4\pi}\,n(R,r_s,t_{\rm act}),
\ee
if the intensity is isotropic. However, in all cases of interest for us,
UHECRs do not propagate fully diffusively and thus we have to supply
information on the angular dependence of the UHECR intensity. It has been
shown in Ref.~\cite{Harari+2016} that a Fisher distribution provides a
good description of the angular distribution of the arrival
directions of UHECRs after stochastic deflections in the EGMF. Hence, the resulting distribution of arrival directions $\theta_{\rm f}$ with respect to the source direction is computed from a random value $r_{\rm u}$ choosen from a uniform distribution over $[0,\,1)$ according to 
\be
\theta_{\rm f}=\arccos{\left\{1+\ln{\left(1-r_{\rm u}[1-\exp(-2\kappa)]\right)}/\kappa\right\}}.
\label{eq:FisherDistr}
\ee
From numerical simulations
of UHECR sources with a finite lifetime the concentration parameter $\kappa$
that characterizes the Fisher distribution can be estimated
by~\cite{Harari+2021}
\be
\kappa \simeq \frac{l_{\rm D}}{r_s}\,\left[ 2+\exp\left(-\frac{2r_s}{3l_{\rm D}}-\frac{1}{2}\left(\frac{r_s}{l_{\rm D}}\right)^2 \right) \right]+\frac{0.44}{[ct_{\rm act}/r_s+1]^{0.8+0.4l_{\rm D}/(ct_{\rm act}+r_s)}-1} .
\label{eq:kappa}
\ee
As these simulations do not account for the change of rigidity we will
subsequently use the average concentration parameter
\be
\bar{\kappa}(R, r_s,t_{\rm act}) \equiv \langle \kappa(R_0(R,x), r_s,t_{\rm act}) \rangle_x = \frac{1}{r_s}\int_0^{r_s} \diff x\,\,\kappa(R_0(R,x), r_s,t_{\rm act}) ,
\label{eq:mean_kappa}
\ee
similar to the average enhancement factor~(\ref{eq:mean_enh}) that is used to
correct the absolute value of the flux. Here, the second term of
Eq.~(\ref{eq:kappa}) takes into account the finite lifetime of the source.
The UHECR arrival maps that account for a turbulent and isotropic EGMF are subsequently determined using Eq.~(\ref{eq:FisherDistr}) for a given $\bar{\kappa}$ under consideration of the limited statistics of the most recent anisotropy study \cite{deAlmeida+2021_ICRC} of the Auger experiment. Moreover, we include the deflections by the JF12 Galactic magnetic field model using the so-called lensing technique, where anti-particles have been propagated backwards to obtain the trajectories of the regular particles that hit the Earth \cite{Bretz+2014}. 

\begin{figure}[htbp]
\centering
\includegraphics[width=.355\linewidth]{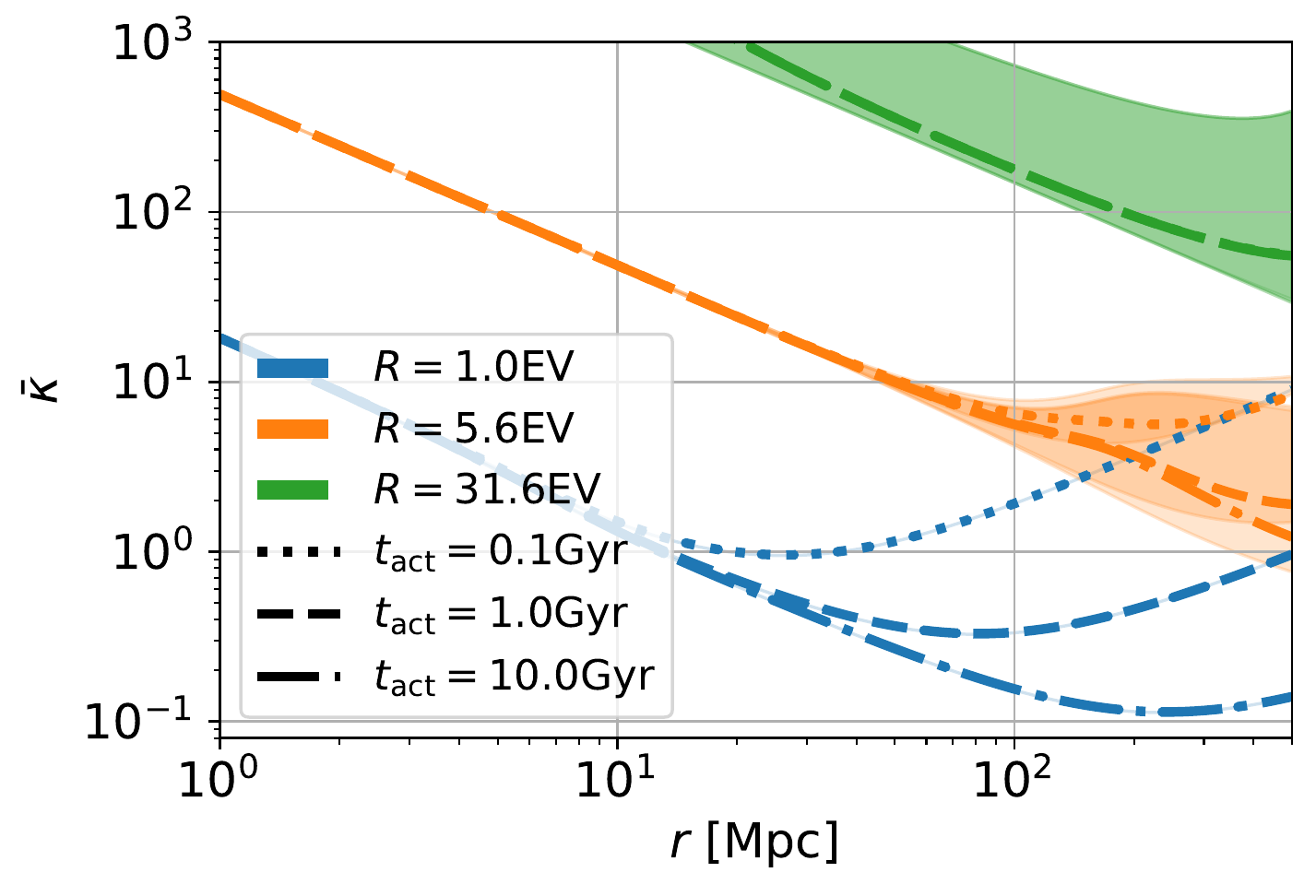}
\includegraphics[width=.595\linewidth]{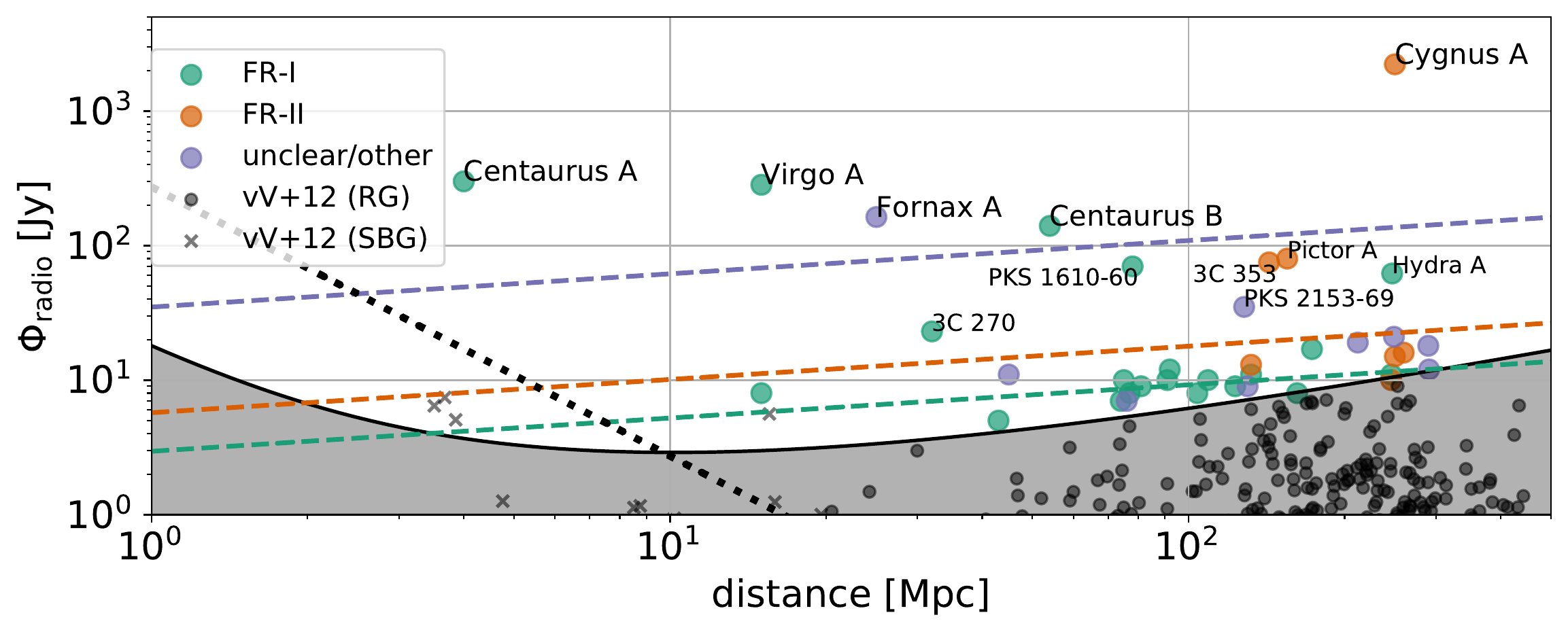} 
\caption{\emph{Left}: The mean concentration parameter~(\ref{eq:mean_kappa}) as function of the source distance for three different source lifetimes and rigidities. The edges of the coloured bands indicate the corresponding concentration parameter~(\ref{eq:kappa}) for $R$ and $R_0(R,r)$, respectively.
  \emph{Right}: The radio flux of the considered source sample~\cite{RachenEichmann2019_catalog} (colored dots), as well as the dim sources (black dots and crosses) from the catalog of van Velzen et al.~\cite{vanVelzen+2012} that are (except for three close-by starburst galaxies) below the conservative source selection criterion \cite{RachenEichmann2019_catalog} --- as indicated by the black solid line. Above the black dotted line sources are likely able to provide a maximal rigidity $\hat{R}>1\,\text{EV}$ and above the colored dashed lines their CR flux is in case of pure rectilinear propagation at least $20\,\%$ (purple), $5\,\%$ (orange), or $2\,\%$ (green) of the brightest CR source in our sample (Cen A). The exact location of these constraints depends on several parameters. Here, the constraints are shown for an optimistic scenario using $\beta_L=0.89$, $g_{\rm m}=4/7$, $g_{\rm acc}=1$, and $k=0$.
}
\label{fig:kappa_sources}
\end{figure}

In case of high rigidities (see the green lines in the left panel of
Fig.~\ref{fig:kappa_sources}), where $l_{\rm D}\gg r_s$, the impact of the
finite lifetime vanishes, since CRs propagate quasi-rectilinearly. In this
case, we read from Fig.~\ref{fig:kappa_sources} that $\kappa\gg 10$  which
corresponds to a mean deflection $\ll 23\degree$ with respect to the source
direction. Hence, the change of rigidity can have a significant effect on
$\kappa$. However, only at intermediate rigidities
(see the orange lines in the left panel of Fig.~\ref{fig:kappa_sources}),
where $l_{\rm D}\sim r_s$ and $t_{\rm act}c \gtrsim r_s$,
this leads to a significant change of the angular distribution of CR
which corresponds to $0.1\lesssim\kappa\lesssim 10$.
Now $\bar{\kappa}$ is typically close to
$\kappa(R)$ indicating that the CR propagates most of its distance with a
rigidity that is rather close to the observed one. 

For the choice of local radio galaxies included in our analysis, we
use the catalogue defined in Ref.~\cite{RachenEichmann2019_catalog}. 
This catalogue is based on the source catalog of van Velzen et
al.~\cite{vanVelzen+2012}  which used the 2MRS catalog that selects
sources by their infrared flux. In addition, it contains  several
powerful radio sources that have previously been missed by van Velzen et
al. Moreover, only local sources up to 500\,Mpc distance to Earth are
considered, because at greater distances the UHECR flux is surpressed
significantly, either by attenuation (via $\bar{\eta}$ at
$R\gtrsim 10\,\text{EV}$) or by diffusion effects (via $\bar{\xi}$ at
$R\lesssim 1\,\text{EV}$).

In addition, there is a low probability that sources at such large distances
provide an individual flux contribution that is comparable to the one from
nearby sources (as indicated by the dashed colored lines in the right panel
of Fig.~\ref{fig:kappa_sources}). On the other hand, those very nearby radio
sources with a radio flux of a few~Jy (which are predominantly starburst
galaxies) can hardly provide the necessary energies (as indicated by the
black dotted line in the right panel of Fig.~\ref{fig:kappa_sources}). 
Thus, the complete local source sample consists of 39~sources, whereof only
seven can be clearly identified as FR-II type. We have verified that only
in the case of the compact object 3C\,111 a moderate boosting is justified.
For these seven FR-II sources dynamical evolution models provide a fairly
accurate estimate on their jet power as well as their jet lifetime, which
are shown in Table~\ref{tbl:RGjetData}. 

\begin{table}[h!]
\begin{minipage}{\linewidth}
\centering
\caption{Individual jet characteristics.}
  \begin{tabular}{ l c c c c}
  \toprule
            Source name  & FR type &$Q_{\rm jet}$ [$10^{44}$\,erg/s] & $t_{\rm act}$ [Myr] &  Reference \\
   \midrule
      3C\,33 & II & $2$ & $61$ & \cite{Machalski+2021}   \\ 
     3C\,98 & II & $7$ & $5.5$ & \cite{Machalski+2021}  \\
    3C\,353 & II & $2.3$ & $22.1$ & \cite{Machalski+2021}  \\
    3C\,390.3 & II & $1.2$ & $85$ & \cite{Machalski+2021}  \\
    3C\,430 & II & $0.68$ & $23$ & $^{[a]}$  \\
    Pictor\,A & II & $2.4$ & $61$ & $^{[a]}$  \\
    Cygnus\,A & II & $83$ & $7$ & \cite{GodfreyShabala2013,Machalski+2007,Carilli+1991}, $^{[a]}$ \\
    Centaurus\,A & I & $0.1$ & $100$ & \cite{Neff+2015,Eilek2014,Wykes+2013,Croston+2009} \\
   \bottomrule 
\end{tabular}
\label{tbl:RGjetData}
\end{minipage}
\centering
\footnotesize{$^{[a]}$ Private communication with Jerzy Machalski (calculations are based on their recent model~\cite{Machalski+2021}).}
\end{table}

\begin{figure}[htbp]
\centering
\includegraphics[width=.495\linewidth]{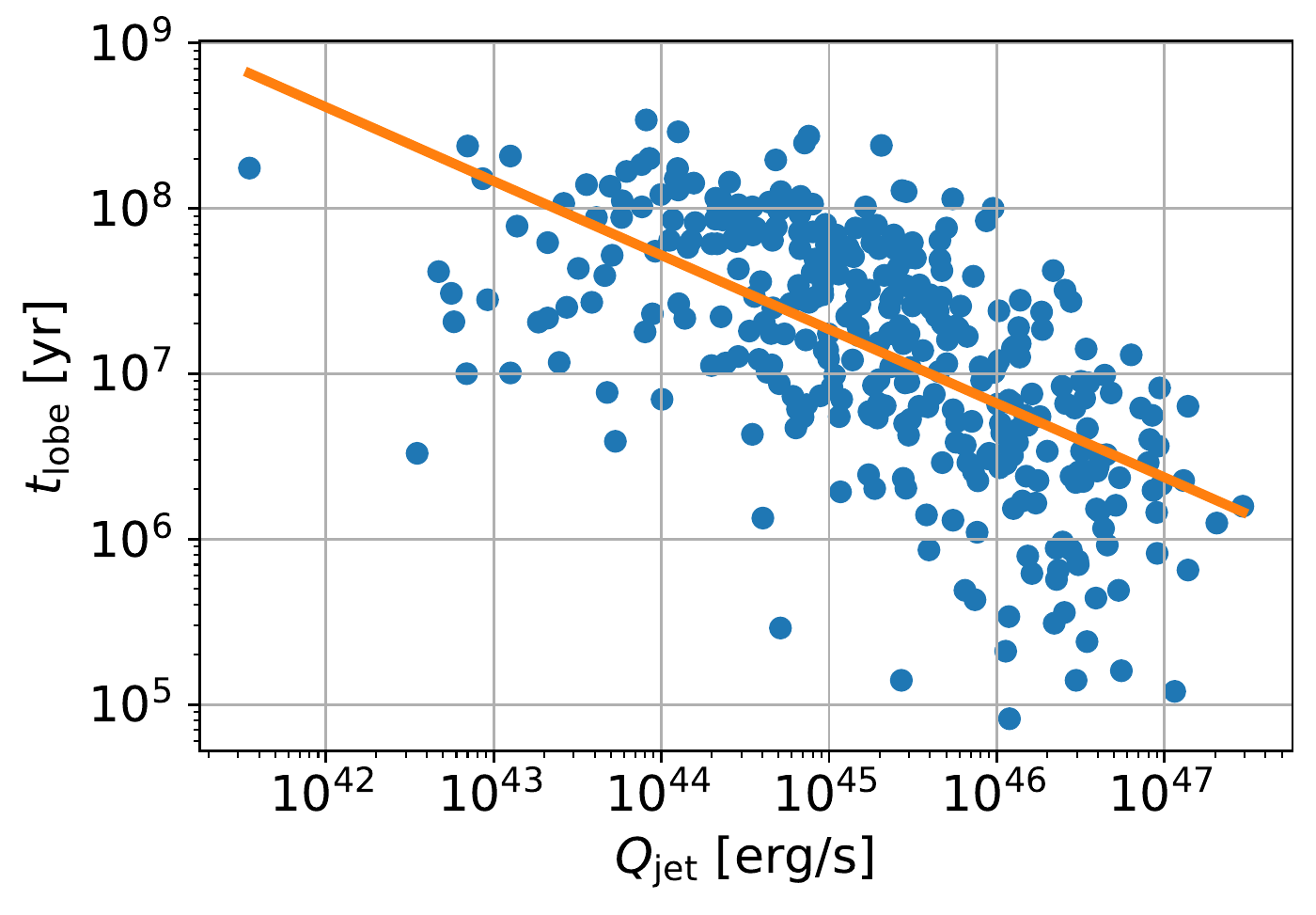}
\caption{The lobe age of 361~FR-II sources dependent on their jet power (taken from \cite{Machalski+2021}), as well as a heuristic correlation function as given by Eq.~(\ref{eq:lobeAge-jetPower-corr}).
}
\label{fig:lobeAge-jetPower-corr}
\end{figure}

For the other sources, only in case of Centaurus~A an individual assignment of
its jet properties can be performed. However, even in this case these values
are only an order of magnitude estimate. For the other sources, we apply the
general radio-jet power correlation~(\ref{Qcr}). The source lifetime
$t_{\rm act}$ is treated as a free parameter for all FR-I sources except
Centaurus~A
as well as in the case of those radio galaxies of unclear/other type with
a radio brightness below $2\times 10^{25}\,\text{W}\,\text{Hz}^{-1}\,\text{sr}^{-1}$ at 151\,MHz---which corresponds to the common transition power
between FRs sources of type~I and type~II~\cite{FanaroffRiley1974}.
In the case of more powerful sources of unclear/other type we adopt the
dynamical evolution model from Ref.~\cite{Machalski+2021} for FR-II sources
and estimate the heuristic correlation function 
\be
t_{\rm lobe} = 52\,\text{Myr}\,(Q_{\rm jet}/10^{44}\,\text{erg/s})^{-0.45}
\label{eq:lobeAge-jetPower-corr}
\ee
between the jet power $Q_{\rm jet}$ and the age of the lobes $t_{\rm lobe}$
shown in Fig.~\ref{fig:lobeAge-jetPower-corr}. 

Individual sources can deviate from this correlation by more than an
order of magnitude---in particular those that do not possess a typical
FR-II morphology.
Still, using this correlation is more justified than employing the same
lifetime for all sources. Furthermore, it has also been confirmed by other
observations~\cite{TurnerShabala2015} that the lower the jet power, the greater
the chance of a long lifetime of high-luminosity radio galaxies. However, in
case of low-luminosity radio galaxies (which have in their vast majority a FR-I
morphology) this inverse correlation does not hold anymore as shown e.g.\ by
Turner and Shabala~\cite{TurnerShabala2015}. Therefore we treat $t_{\rm act}$ as
a single free parameter, common to all low-luminosity radio sources. For the
others, i.e.\ the high-luminosity radio sources, we will subsequently suppose
that their lifetime can be approximated by the age of their lobes, i.e.\
$t_{\rm act}\simeq t_{\rm lobe}$, either given in Table~\ref{tbl:RGjetData}
or by Eq.~(\ref{eq:lobeAge-jetPower-corr}).

\subsection{UHECRs from the bulk of radio galaxies}
\label{sec:Bulk}

In addition to the UHECRs from these bright, individual local radio galaxies, which we suppose to dominate the contribution on local scales, there is a
diffuse contribution from the bulk of distant radio galaxies. 
Studies of the Local Supercluster indicate \cite{Peebles2021} that at redshift $\gtrsim 0.02$ inhomogenities in the large-scale distribution
of the sources vanish and in case of a strong EGMF they will already be averaged out on smaller scales, so that we can assume this contribution to be isotropic. 
Moreover, these sources fill space nearly homogeneously with
UHECRs, if the particles are able to propagate at least about the mean distance
between the considered source class. According to the propagation
theorem~\cite{AloisioBerezinsky2004}, the diffuse energy spectrum is 
independent of the particle propagation mode if the spatial separation between
sources is smaller than all other characteristic length scales of propagation.
Hence, we expect no modification of the diffuse spectrum by the EGMF
and the enhancement factor $\bar \xi$ can be neglected.

Based on the radio luminosity function of low- and high-luminosity radio
sources from Ref.~\cite{Willott+2001}, we use the so-called continuous
source function (CSF) 
\begin{equation}
  \Psi_{0,i}(R,\,z) \equiv
  \frac{\diff N_{\rm cr}(Z_i) }{ \diff V \mathrm{d}R\,\diff t} = 
\int 
 S_{0,i}\big(R,\hat R(Q_{\rm cr})\big)\,\frac{\diff N }{
\diff V\,\diff Q_{\rm cr}}\,\diff Q_{\rm cr} 
\label{CRsourceRateDensity}
\end{equation}
following the approach of Refs.~\cite{Eichmann2019,Eichmann2019ICRC}. 
Here, $\diff N /(\diff V\,\diff Q_{\rm cr})$ is the number of radio sources
per volume and luminosity, while $S_{0,i}(R,\hat R(Q_{\rm cr}))$ is the
UHECR injection rate of a radio source with the maximal rigidity
$\hat  R(Q_{\rm cr})$.
Using a sharp  cutoff at $\hat R(Q_{\rm cr})$ instead of an exponential
suppression, Eq.~(\ref{CRsourceRateDensity}) can be solved analytically, for
the solution see Eq.~(3.17) in Ref.~\cite{Eichmann2019}. 
To account for the impact of the CMB and the EBL, we employ again the mean modification factor as given by
\be
\bar{\eta}_{\rm csf}(R,z) = \langle \eta_{{\rm csf},i}(R,z) / A_i \rangle_i  =
\left\langle \frac{1}{A_i}\,\frac{\Psi_{i}(R,z)}{\Psi_{0,i}(R,z)} \right\rangle_i.
\ee
For the calculation of this function, we perform again a large set of 1D
CRPropa simulations and compute the ratio $\Psi_{i}(R,z)/\Psi_{0,i}(R,z)$
of the attenuated and the unattenuated CSFs at a given redshift $z$.
Because of the integration over the CR power $Q_{\rm cr}$, the modification
factor $\bar{\eta}_{\rm csf}(R,z)$ depends not only on the assumed spectral
index $\alpha$ but also on the parameters $g_{\rm acc}$, $g_{\rm m}$, $k$
and $\beta_{\rm L}$ that define the correlation between radio and CR power.

\begin{figure}[htbp]
\centering
\includegraphics[width=.495\linewidth]{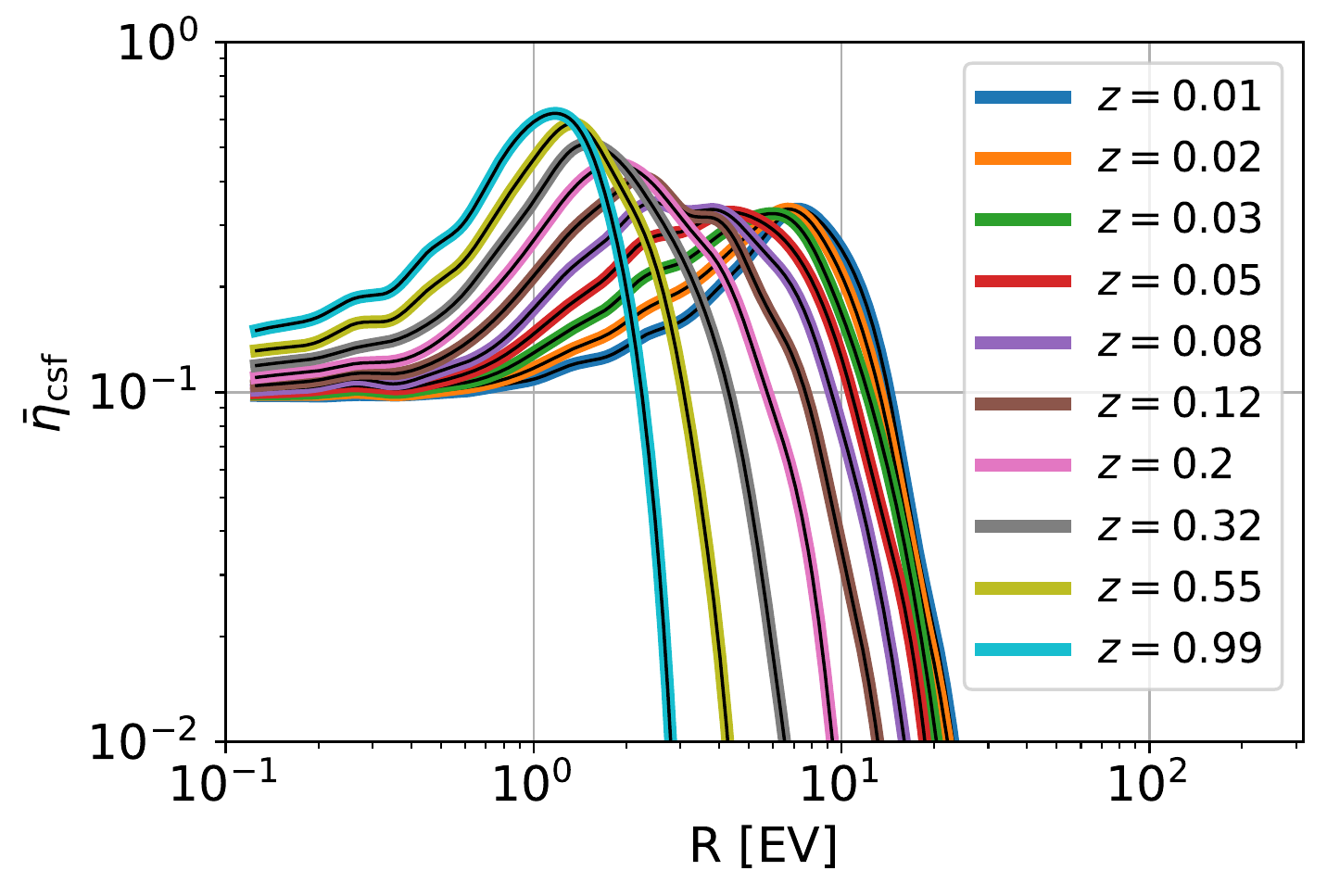}
\includegraphics[width=.495\linewidth]{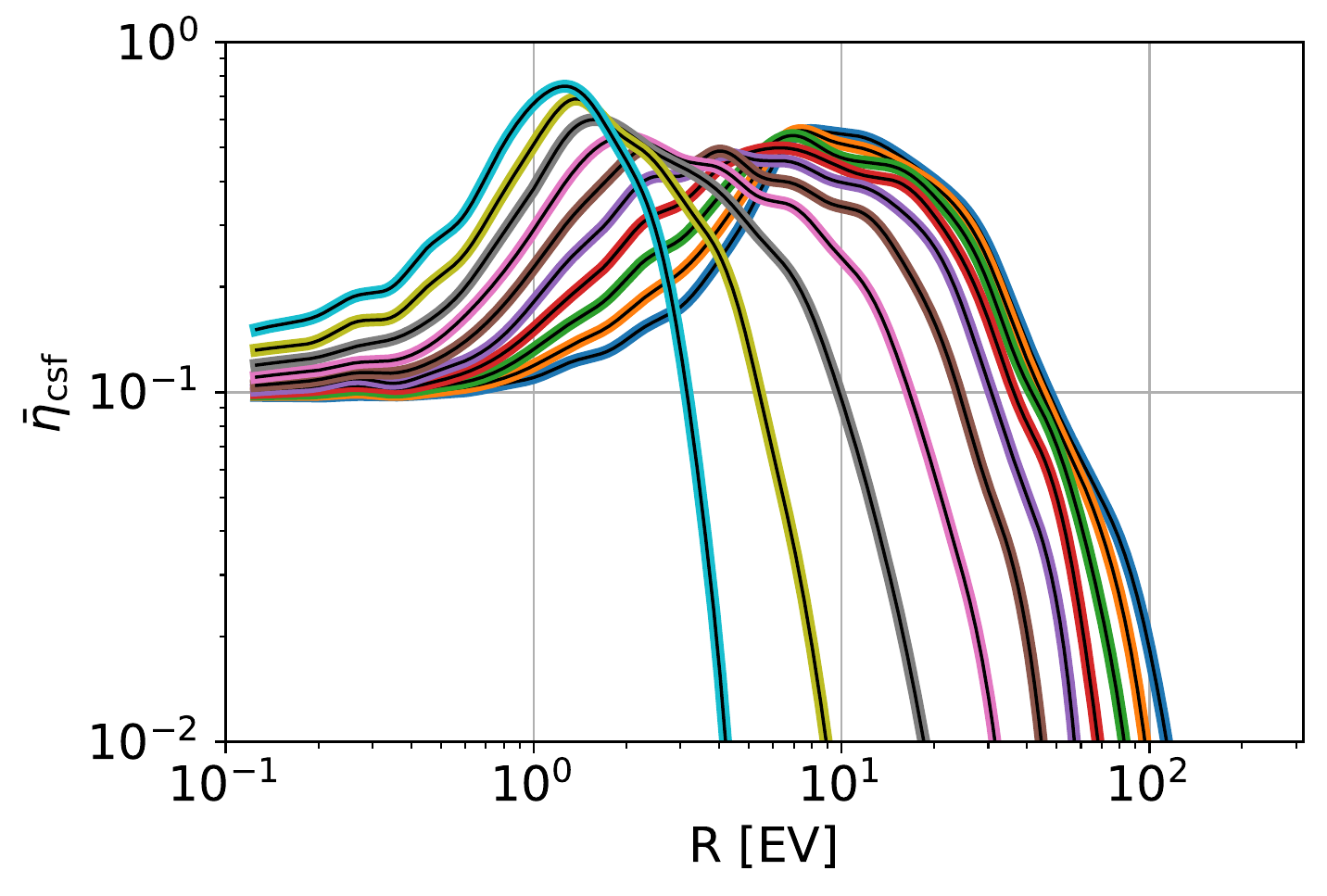} 
\caption{The average modification factor $\bar{\eta}_{\rm csf}$ for the bulk of low-luminosity (\emph{left}) and high-luminosity (\emph{right}) radio sources, respectively, with $\alpha=2$, $g_{\rm acc}=0.5$, $g_{\rm m}=0.6$, and $k=0.5$.
}
\label{fig:modFactor_CSF}
\end{figure}

The deviations between $\eta_{{\rm csf},i}(R,z) / A_i$ for different nuclei are
in general smaller than those shown in the right panel of
Fig.~\ref{fig:interactionEffects}, because of the larger source distances.
The difference of $\bar{\eta}_{\rm csf}(R,z)$ between low- and high-luminosity
radio sources shown in Fig.~\ref{fig:modFactor_CSF} exposes  that the latter
provide a larger amount of secondaries, as they are more numerous at high
redshifts than low-luminosity radio sources and are in principle able to
accelerate UHECRs to higher rigidities.  The total contribution to the
UHECR intensity by the bulk of radio galaxies between a redshift $\check{z}$ and $\hat{z}$ is given by
\be
J_{\rm csf}(R,t_{\rm act})=\frac{c}{4\pi}\,\int_{\check{z}}^{\hat{z}}
\diff z\,\,\left| \frac{\diff t}{\diff z} \right| \sum_i \Psi_{0,i}(R,\,z)\,A_i\,\bar{\eta}_{\rm csf}(R,z)
\ee
with
\be
\left| \frac{\diff t}{\diff z} \right| = \left[ H_0\,(1+z)\,\sqrt{(1+z)^3\Omega_{\rm m}+\Omega_\Lambda} \right]^{-1}\,,
\ee
where we use the standard $\Lambda$CDM cosmology with a Hubble constant
$H_0=70\,\text{km}\,\text{s}^{-1}\,\text{Mpc}^{-1}$, $\Omega_{\rm m}=0.3$,
and $\Omega_\Lambda=0.7$.  Furthermore, we obtain that
$\sum_i A_i \Psi_{0,i}(R,\,z)\simeq 2\, \Psi_0(R,\,z)$, with $\Psi_0(R,\,z)$
as the total initial CSF, is again \emph{independent of the initial
  elemental abundances}, since the mean charge number
$\bar{Z}\equiv \sum_i f_i\,Z_i$ (that is used in Eq.~(3.17) of Ref.~\cite{Eichmann2019}), where $f_i$ denotes the fractional abundances of a given
element $i$, and $A_i\simeq 2\,Z_i$.

\section{Results and comparison to data}
\label{sec:FitResults}

The goal of this work is to investigate if a sample of individual local
radio galaxies is sufficient to explain the mean features of the UHECR
data, and to which amount a diffuse contribution by the bulk of radio
galaxies is in addition needed. Further, if such a sample exists we try
to identify the most relevant individual sources and their characteristics,
in particular the lifetime of low-luminosity sources, their CR power and
the maximal rigidity.

\subsection{From rigidity to energy at Earth}

\begin{figure}[htbp]
\centering
\includegraphics[width=.495\linewidth]{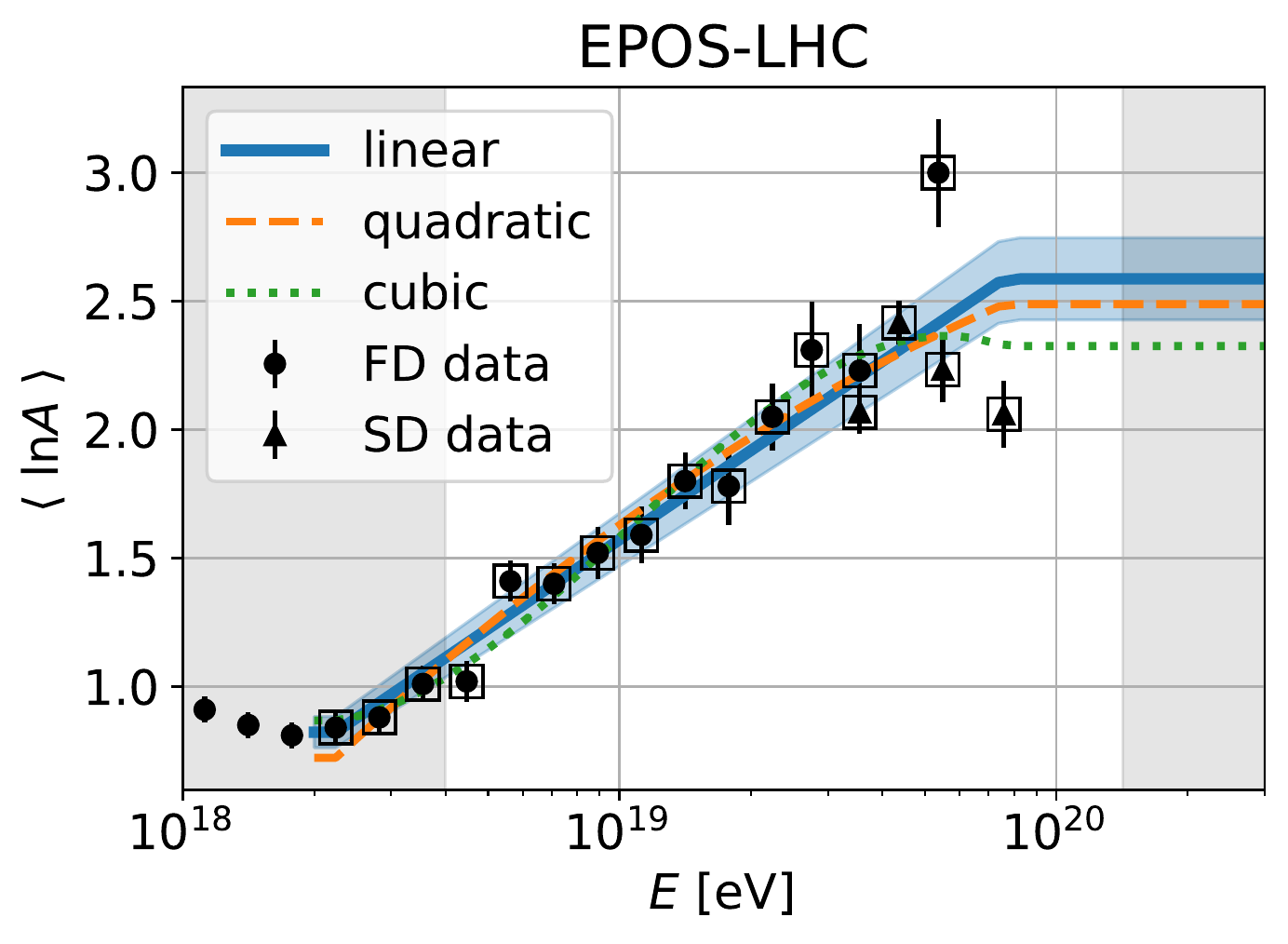} 
\includegraphics[width=.495\linewidth]{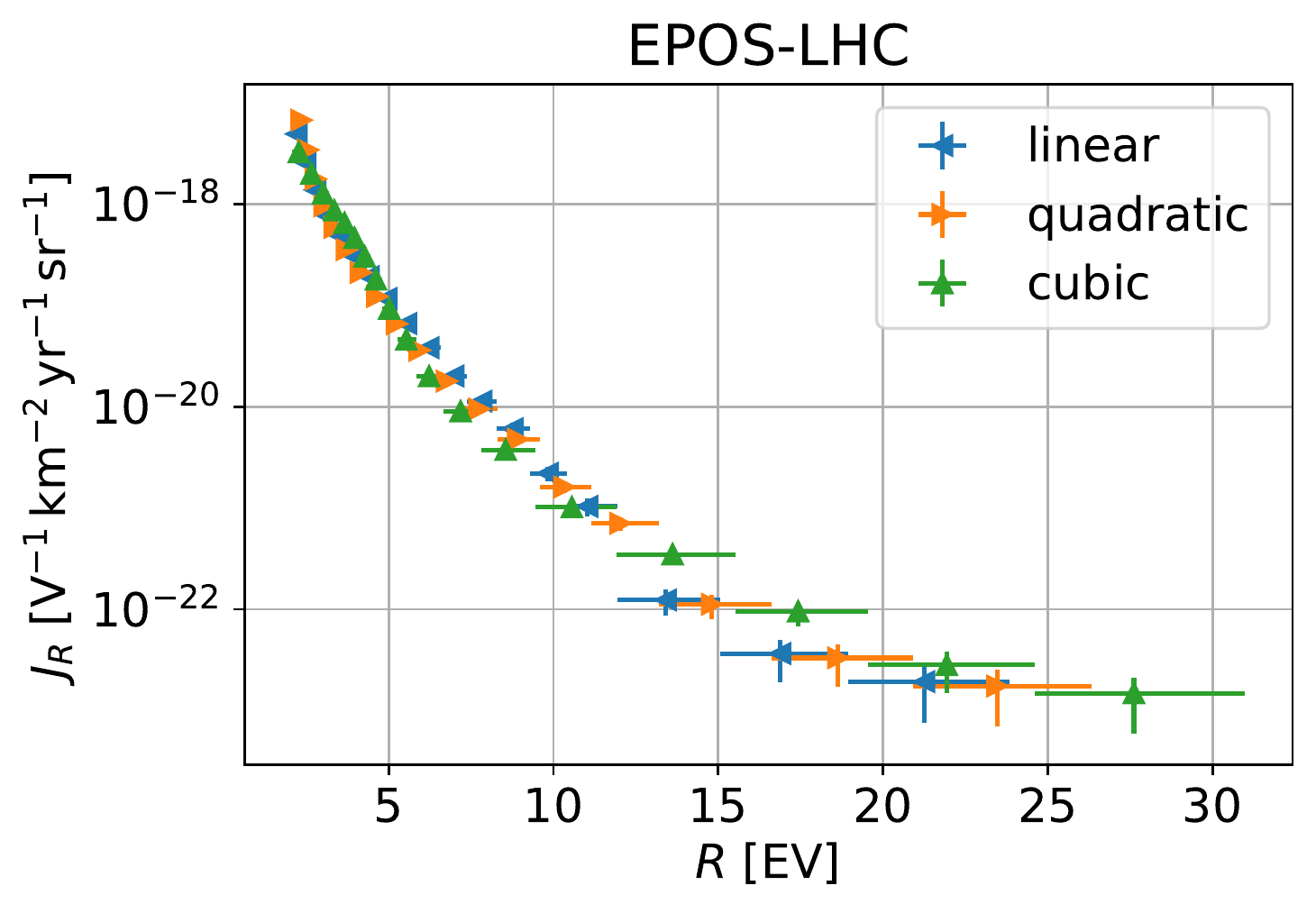}
\caption{
Using the composition data for the hadronic interaction model EPOS-LHC to convert rigidity into energy.
\emph{Left}: Inferred mean logarithmic mass of UHECRs $\langle \ln A\rangle_{\rm obs}$ as a function of energy as well as three different splines (polynomials of different degrees) that interpolate those data that is marked by an open square. In case of the linear spline (blue solid line) also the uncertainty range is shown. The greyed area marks the energy range that is not considered in our analysis. The inferred $\langle \ln A\rangle_{\rm obs}$ values are based on measurements of the depth of shower maximum with the Auger fluorescence detectors (FD)~\cite{Yushkov:2020nhr} and surface detectors (SD)~\cite{PierreAuger:2017tlx}. 
\emph{Right}: The conversion of the Auger flux $J_E$ into $J_R$ using the different splines of $\langle \ln A\rangle_{\rm obs}$ that are illustrated in the left panel.}
\label{fig:lnAdata-rigFlux}
\end{figure}

To compare our model with data, the calculated UHECR intensity $J(R)$ needs
to be converted into a function of energy.  The total CR intensity
$J_R=J_{\rm csf}+\sum_s J_s$ at Earth is composed of the sum of the diffuse
flux from low- and high-luminosity radio galaxies as well as the local sample
of radio-bright sources. For the latter we will investigate three different sub-samples (consisting of 5, 11 and 26 sources respectively), that are selected according to their CR flux based on the CR power $Q_{\rm CR}$ defined in
Eq.~(\ref{Qcr}) assuming rectilinear propagation (see the right panel of
Fig.~\ref{fig:kappa_sources}). Based on the supposed isotropy of the CSF contribution, we subsequently include only sources beyond the Local Supercluster, choosing $\check{z}=0.02$. However, our results hardly depend on this parameter choice because the bulk of radio galaxies at $z<0.02$ predominantly affects the total CSF contribution at high rigidities (above about 10\,EV), where the individual local sources are expected to dominate the UHECR flux. At $z>1$ the CSF contribution above a few EV already vanishes, as can be seen in Fig.~\ref{fig:modFactor_CSF}, so that without loss of generality we use $\hat{z}=1.5$ in the following. Since $J_R$ does not depend on the unknown
initial elemental abundances at the sources, we avoid a multitude of
additional free parameters. We convert the total rigidity spectrum at Earth
into an energy spectrum using the observed mean logarithm of the mass number,
$\langle \ln A\rangle_{\rm obs}$. In doing so, we do not account for the
variance of the $\ln A$ distribution. This information could be used to
constrain the initial CR composition at the sources, which is however beyond
the scope
of this work. Nevertheless, we like to stress that our approach does not
rely on a pure composition at Earth, i.e. the $\langle \ln A\rangle_{\rm obs}$
contribution of the individual sources at Earth can differ, since either a
single source or even multiple sources can contribute different CR nuclei at
a given energy $E_{\rm obs}$ which in total sum up to the observed
compositional data.\footnote{As a simple illustrative example: Imagine a 50/50 mixing of a light (l) and a heavy (h) element at the same energy $E_{\rm obs}$, which corresponds to the rigidities $R_{\rm obs}^{\rm (l)}>R_{\rm obs}^{\rm (h)}$. Converting these rigidities to the initial rigidities at the sources this difference increases dependent on the source distance $d$, so that at large distances and high rigidities the corresponding source rigidities are $R_{0}^{\rm (l)}(d)\gg R_{0}^{\rm (h)}(d)$. Note that $R_{0}=E_{0}/Z_{0}$ and one could either set the energy or the charge number of the CR to adjust the source ejecta at the given rigidity. The most naive approach to realize the requested composition at a given energy would be to use only a single source and a single initial, heavy CR species at two different initial energies and adjust the spectral index so that the requested 50/50 mixing of light and heavy elements is obtained at Earth. However, this approach will likely contradict the observed spectral behavior of the CRs at Earth. Hence, in case this composition mixing shall be explained by a single source with a fixed spectral index, one needs to adjust the relative initial fractions of two different CR species in such a manner that their resulting flux at Earth is equal. In case of multiple sources, one might alternatively also obtain the requested 50/50 mixing by the \emph{same\/} CR species emitted by different sources (most likely at different energies). }
Hence, we suppose that the observed rigidity can be estimated by
$R_{\rm obs}(E,Z)=E_{\rm obs}/(eZ_{\rm obs})$, with
$Z_{\rm obs}=\exp(\langle \ln A\rangle_{\rm obs})/2$. This is 
a good approximation for all nuclei except for protons which can be neglected
in the energy range of interest for us. The conversion of the CR intensity
from rigidity to energy is given by
\be
J_E(E) = J_R(R)\,\biggl| \frac{\diff R_{\rm obs}}{\diff E_{\rm obs}} \biggl|
= \frac{J_R(R)}{e}\,\left[ \frac{1}{Z_{\rm obs}} - \frac{E_{\rm obs}}{Z_{\rm obs}^2}\,\frac{\diff Z_{\rm obs}}{\diff E_{\rm obs}} \right]\,,
\label{spectrum-conversion}
\ee
and the  corresponding standard deviation by
\be
\Delta J_E(E) = \frac{J_R(R)}{e}\,\sqrt{ \left[ \frac{2E_{\rm obs}}{Z_{\rm obs}^3}\,\frac{\diff Z_{\rm obs}}{\diff E_{\rm obs}} - \frac{1}{Z_{\rm obs}^2} \right]^2\,(\Delta Z_{\rm obs})^2
+ \frac{E_{\rm obs}^2}{Z_{\rm obs}^4}\, \left(\Delta \frac{\diff Z_{\rm obs}}{\diff E_{\rm obs}} \right)^2 }
\ee
with
\be
\Delta Z_{\rm obs} = Z_{\rm obs}\,\Delta\langle\ln A\rangle_{\rm obs}\qquad\text{and}\qquad 
\Delta \frac{\diff Z_{\rm obs}}{\diff E_{\rm obs}} = Z_{\rm obs}\,\frac{\diff \langle \ln A \rangle_{\rm obs}}{\diff E_{\rm obs}}\,\Delta\langle\ln A\rangle_{\rm obs}\,. 
\ee
This flux conversion demands a non-vanishing derivative
$\diff R_{\rm obs}/\diff E_{\rm obs}$ and therefore we have to neglect the
irregularities visible in the composition data shown in the left panel of
Fig.~\ref{fig:lnAdata-rigFlux}. Determining $Z_{\rm obs}(E_{\rm obs})$ by a
linear spline function between 2\,EeV and 100\,EeV yields a quite accurate
description of the $\langle \ln A\rangle_{\rm obs}$ data, in particular at
lower energies. 
The fit using a higher order polynomial shown in addition does not provide a
significantly better approximation of the composition data. Nevertheless,
a change of the fit function will lead to minor changes in $R_{\rm obs}$ as
well as $J_E$. To illustrate these differences, we use the observed energy
spectrum $J_E$ and apply Eq.~(\ref{spectrum-conversion}) to convert it into
the rigidity spectrum $J_R$ shown in the right panel of
Fig.~\ref{fig:lnAdata-rigFlux}. Clearly, the resulting rigidity spectrum
depends on the  composition data and their accuracy, what is, however, a
general issue for every approach that tries to explain the UHECR data.
Characterizing UHECRs from their sources to Earth only by their rigidity
provides the benefit that we can use the rather well-known composition data
at Earth to split the rigidity information into energy and charge/mass
number, instead of introducing an arbitrary source composition in the first
place. Thus, our fitting approach ensures a priori that the
$\langle \ln A\rangle_{\rm obs}$ data are described with high precision
according to the linear spline shown in the left panel of
Fig.~\ref{fig:lnAdata-rigFlux}.
At high energies, we include the composition data obtained with the Auger
surface detectors~\cite{PierreAuger:2017tlx}, to obtain a better constraint
on $Z_{\rm obs}$ at and beyond the last data point derived from the
fluorescence detector data~\cite{Yushkov:2020nhr}. Unfortunately,
these data indicate opposite trends. 

Unless otherwise stated, we will use the $\langle \ln A\rangle_{\rm obs}$
prediction based on the hadronic interaction model
EPOS-LHC~\cite{PhysRevLett.101.171101,PhysRevC.92.034906}, as the resulting
mean mass numbers are in-between those from the two alternative
interaction models QGSJet~II-04~\cite{Ostapchenko:2010vb,Ostapchenko:2013pia}
and Sibyll2.3c~\cite{Riehn2019_sibyll2.3c}.

\subsection{Fit parameters and procedure}

In what follows, we will compare our model to the observed dipole
characteristics, i.e.\ its strength and direction \cite{deAlmeida+2021_ICRC}, as well as the energy spectrum \cite{PierreAuger:2020qqz} above $5\,\text{EeV}$ from the Pierre Auger
Observatory.\footnote{The $\langle \ln A\rangle_{\rm obs}$ data are implicitly
  fitted as previously described.}
We apply a so-called Galactic lens (more details on the lens and its generation
can be found in Ref.~\cite{Bretz+2014,EichmannWinchen2020}) based on
the JF12 model to account for the CR deflections by the GMF.
Dependent on the rigidity of each UHECR and its arrival direction, the JF12
field leads to deflections up to several tens of degrees of the average arrival
direction at Earth with respect to the source direction. 
The main challenge in a comparison with the UHECR data results from the large
number of dimensions of the parameter space of our model shown in
Table~\ref{tbl:Parameters}. It contains  the source parameters $g_{\rm m}$
and $k$ defining the conversion from jet into CR power, the acceleration
efficiency $g_{\rm acc}$ that sets the maximal rigidity and  the source
spectral index $\alpha$ which depends on the details of the acceleration
process.  Except for $\alpha$, all these parameters are constrained only
within an order of magnitude. In the case of acceleration by a non-relativistic,
strong shock in the outer jet, as induced e.g.\ by its
backflows~\cite{Matthews+2019}, we expect only minor deviations from the
canonical value of $\alpha=2$ leading to $1.5 \lesssim \alpha \lesssim 2.5$,
caused by non-linear effects~\cite{Haggerty+2020, Caprioli+2020}. 
Different acceleration mechanisms including shear acceleration in the
large-scale jet~\cite{1998A&A...335..134O,Rieger:2004jz}, are expected to
lead to smaller values of $\alpha$ and may accelerate particles to
ultra-high energies in (trans-) relativistic jets. 
In addition, we only got some rough constraints on the lifetime of
low-luminosity radio sources~\cite{TurnerShabala2015},
$1\,\text{Myr}\lesssim t_{\rm act}\lesssim 1\,\text{Gyr}$, as well as the
strength of the EGMF~\cite{Pshirkov+2015} $B_{\rm rms}\lesssim 2 \,\text{nG}$
supposing 1\,Mpc as coherence length. In addition, most of these
parameters---in particular $g_{\rm m}$, $k$, $g_{\rm acc}$, and
$t_{\rm act}$---are expected to vary from source to source, because of
differences in the details of their power supply or their evolutionary stage.
Thus, we likely end up with an underdetermined, non-linear system that needs
to be solved. 

Due to limited computational resources, we will only differentiate
$g_{\rm m}$ and $g_{\rm acc}$ between the sources: These two parameters 
impact mainly the spectral relevance of these sources and can incorporate
partly the uncertainty from the radio-jet power correlation.
Finally, we quantify the goodness of the fit by the chi-squared value
$\chi^2(k,\,t_{\rm act},\,B_{\rm rms})=\sum_j (P_j-O_j)^2/\sigma^2(P_j,O_j) $,
where $P_j$ ($O_j$) denotes the model prediction (observation) of the total
diffuse flux as well as the dipole strength and direction at a given energy.
Finally, we use an optimization routine to obtain the global minimum of the
multivariate chi-squared function for a given parameter combination of
$(k,\,t_{\rm act},\,B_{\rm rms})$. Hence, the parameter optimization is only
performed using $g_{\rm m}$, $g_{\rm acc}$ and $\alpha$---which yields in the
case of five individual, local sources (plus the bulk of low- and
high-luminosity radio galaxies) already $15 = 5\times 2+2\times 2 + 1$
parameters that need to be determined within the given constraints. However,
we manage to slightly minimize the parameter space of the optimization
algorithm by using a linear least-squares problem solver to compute
$g_{\rm m}$ (see the Appendix~\ref{app:optProc} for more details).
Table~\ref{tbl:Parameters} summarizes all details on the free (first
three rows) and fixed (last five rows) model parameters used in the fit
procedure.

\begin{table}[h!]
\centering
\small
\caption{Model parameters. The first three rows give the range of the free parameters, while the rest of the parameters are fixed to the quoted values.}
  \begin{tabular}{ l c c c}
  \toprule
            Parameter & Value(s) & Per Source & Description  \\
   \midrule
    $g_{\rm m}$ & $[0.001,\dots,\, 0.9]$ & yes & \small{matter-to-jet power ratio} \\
    $g_{\rm acc}$ & $[0.001,\dots,\, 1]$  & yes & \small{acceleration efficiency}  \\
    $\alpha$ & $[1.5,\dots,\, 2.5]$ & no & \small{source spectral index} \\
    $k$ & $[0.1,\,0.5,\,1,\, 5]$  & no & \small{leptonic-to-hadronic energy density ratio}  \\
    $t_{\rm act}$ [Gyr] & $[0.01,\,0.05,\,0.1,\,0.5,\,1,\,5,\,10]$ & no & \small{low luminosity source lifetime} \\
    $B_{\rm rms}$ [nG] & $[0.1,\,0.5,\,1,\,5]$ & no & \small{rms EGMF strength}  \\
    $l_{\rm c}$ [Mpc]  & 1 & no & \small{EGMF coherence length}  \\ 
    $\check{R}$ [GV]  & 1 & no & \small{minimal CR rigidity}  \\
    $\beta_L$  & 0.89 & no & \small{radio--jet power correlation index} \\
    $\check{z}$  & 0.02 & no & \small{minimal CSF redshift} \\
    $\hat{z}$  & 1.5 & no & \small{maximal CSF redshift} \\
   \bottomrule 
\end{tabular}
\label{tbl:Parameters}
\end{table}

In general, it is quite likely that not only $g_{\rm m}$ and $g_{\rm acc}$, but also $t_{\rm act}$, $\alpha$, and $B_{\rm rms}$ vary (at least slightly) between the sources so that the following fit results will certainly improve by an expansion of the considered parameter space. 
Although, this does not necessarily involve an improved reduced chi-squared result---including the increased number of degrees of freedom. Further, it is not the goal of this work to identify the particular best fit scenario, as this certainly requires more precise data with respect to the Galactic and extragalactic magnetic fields, the source physics as well as the UHECR measurements. 
However, we verified for an individual assignment of $\alpha$ in case of a small local source sample, that the $\chi^2$ results only improve on average by a factor of about 0.9 with no changes on the general outcome, that is shown in the following.

\subsection{Fit results}
\label{sec:fitResults}

\begin{figure}[htbp]
\centering
\includegraphics[width=.99\linewidth]{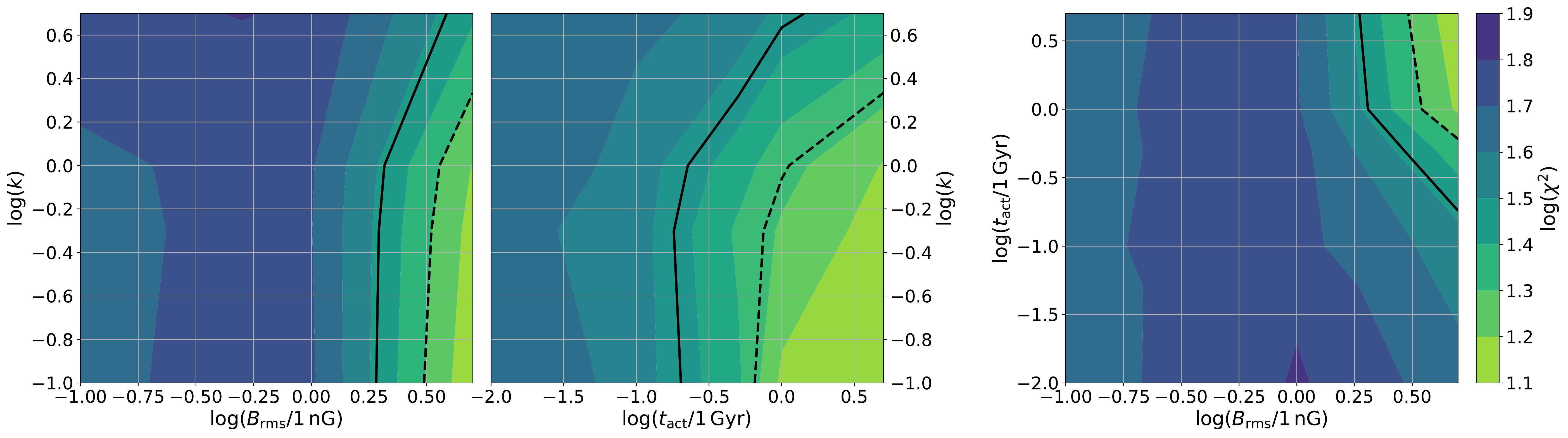} 
\includegraphics[width=.99\linewidth]{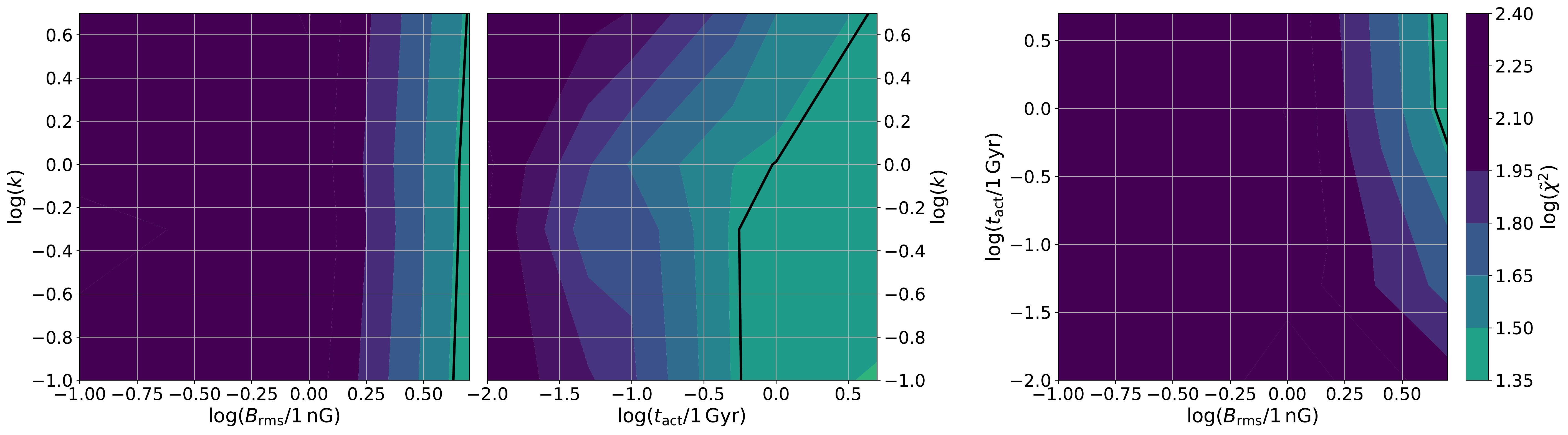} 
\caption{The marginalised (over the third parameter) chi-squared distribution that results from the optimization routine (\emph{upper panel}) as well as after including the residuals by the quadrupole data (\emph{lower panel}). Here the local source sample S5 is used. The black lines correspond to a chi-squared value of $20$ (dashed) and $30$ (solid), respectively.}
\label{fig:distrChi2_S5}
\end{figure}

\begin{figure}[htbp]
\centering
\includegraphics[width=.79\linewidth]{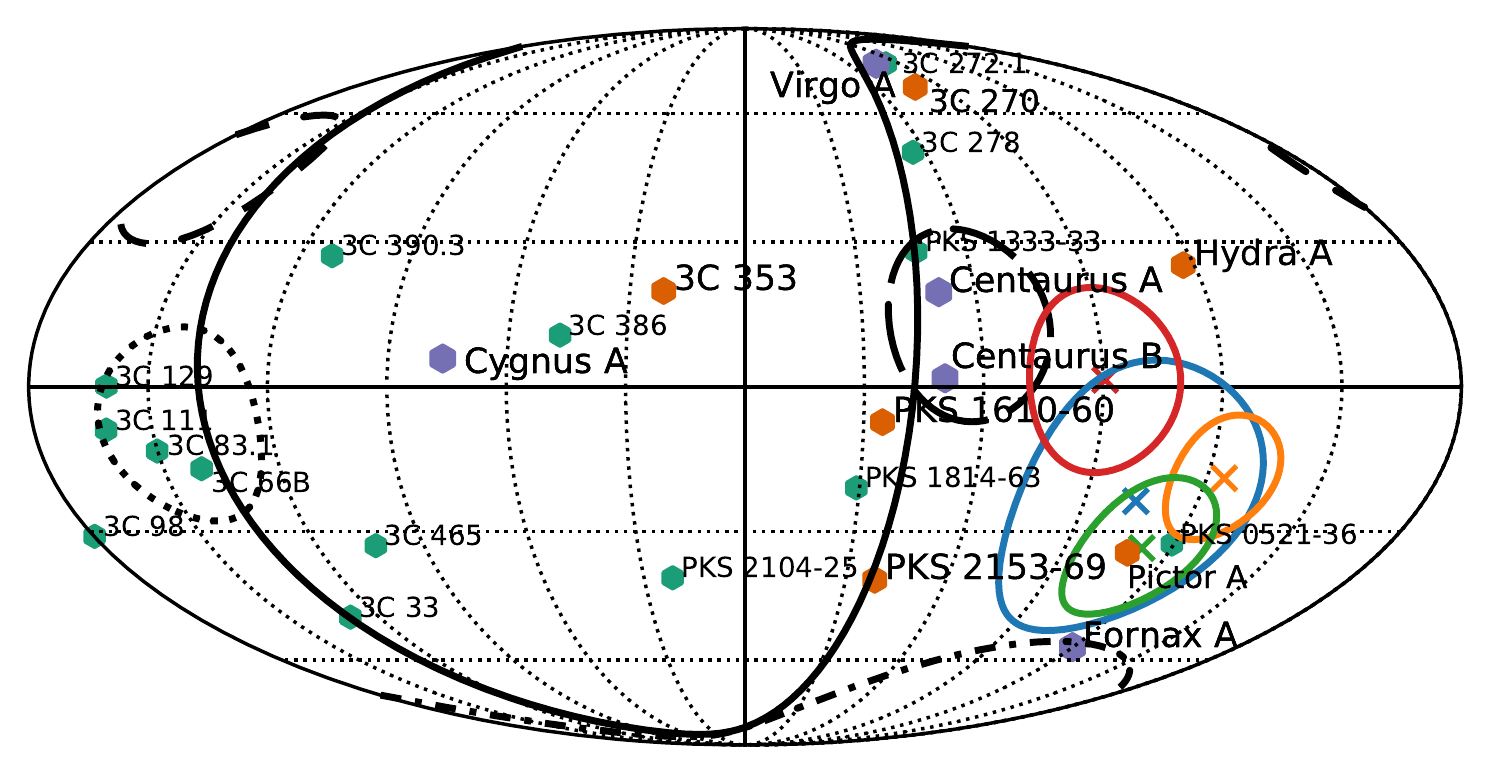} 
\caption{Our sample of radio galaxies shown in Galactic coordinates:
  The S5 sample (purple hexagons), the additional S11 sources (orange hexagons),
  and the additional S26 sources (green hexagons). Colored crosses (and surrounding lines) give the observed dipole direction (and uncertainty) at $[4,\, 8]\,$ EeV (blue), $[8,\, 16]\,$ EeV (yellow), $[16,\, 32]\,$ EeV (green), $>32\,$ EeV (red). The black dashed and dash-dotted lines refer to the three most visible excesses of CR flux above 40 EeV in 2019 as taken from \cite{diMatteo+2019ICRC} and the black dotted line shows the approximate location of the excess that emerged recently \cite{diMatteo+2021ICRC}. The black solid line indicates the Supergalactic plane.}
\label{fig:srcSky}
\end{figure}

The $\chi^2$ results for our fit procedure are shown in the upper panel of
Fig.~\ref{fig:distrChi2_S5} for a local source sample consisting of the 
five brightest CR sources: Centaurus A, Centaurus~B, Virgo~A, Fornax~A, and
Cygnus~A (this smallest sample is subsequently referred to as S5). With such
a small local source sample, the resulting distribution of chi-squared values
exposes that a strong EGMF is needed as well as a rather long lifetime
($\gtrsim 1\,\text{Gyr}$) of Fornax~A and Virgo~A. Including the residuals
with respect to the quadrupole data, this preference becomes stronger without
significantly changing the following outcome: 
First, only Virgo~A and Fornax~A yield a significant contribution of at least
$\sim \text{few}\times 10\%$ to the observed UHECR intensity. Second,
the UHECR data are already explained quite well, apart from the right
ascension of the dipole direction below $32\,\text{EeV}$, which shows
deviations of up to $35\degree$.
The assumed leptonic-to-hadronic energy ratio $k$ has only a minor impact
on the fit. 
The comparison of the different sized local source samples shows that there
is some benefit in increasing S5  by Pictor~A, Hydra~A, 3C\,270, PKS\,1610-60,
3C\,353, and PKS\,2153-69 (hereafter S11). Adding even more sources
does not yield a significant improvement of the resulting $\chi^2$ values.
However, based on the location of these local sources in the sky (see
Fig.~\ref{fig:srcSky}) we do not expect that the small source samples
(S5 or S11) will be able to explain the recent indications of medium-scale
anisotropy~\cite{diMatteo+2019ICRC,diMatteo+2021ICRC}, this issue
will be discussed in see Sect.~\ref{sec:concl}. Moreover, the optimization
algorithm only accounts for the dipole anisotropy which is the most
robust directional information that we currently have. The inclusion of
higher order anisotropies, for which no firm detection exists, would lead to
a significantly higher computational effort.
%

\subsubsection{Eleven local sources}

\begin{figure}[htbp]
\centering
\includegraphics[width=.99\linewidth]{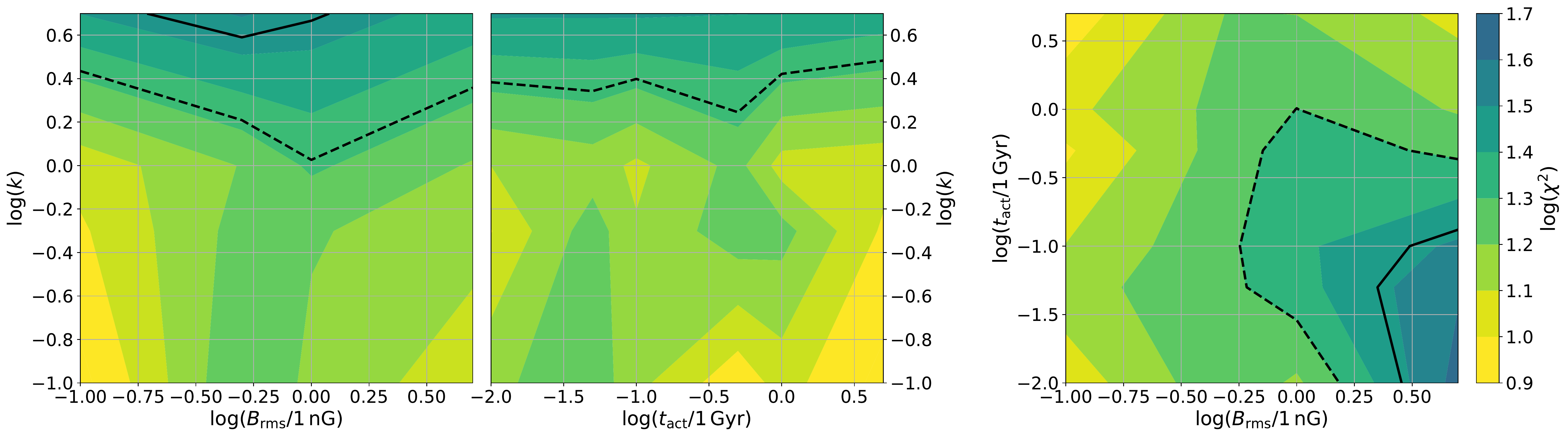} 
\includegraphics[width=.99\linewidth]{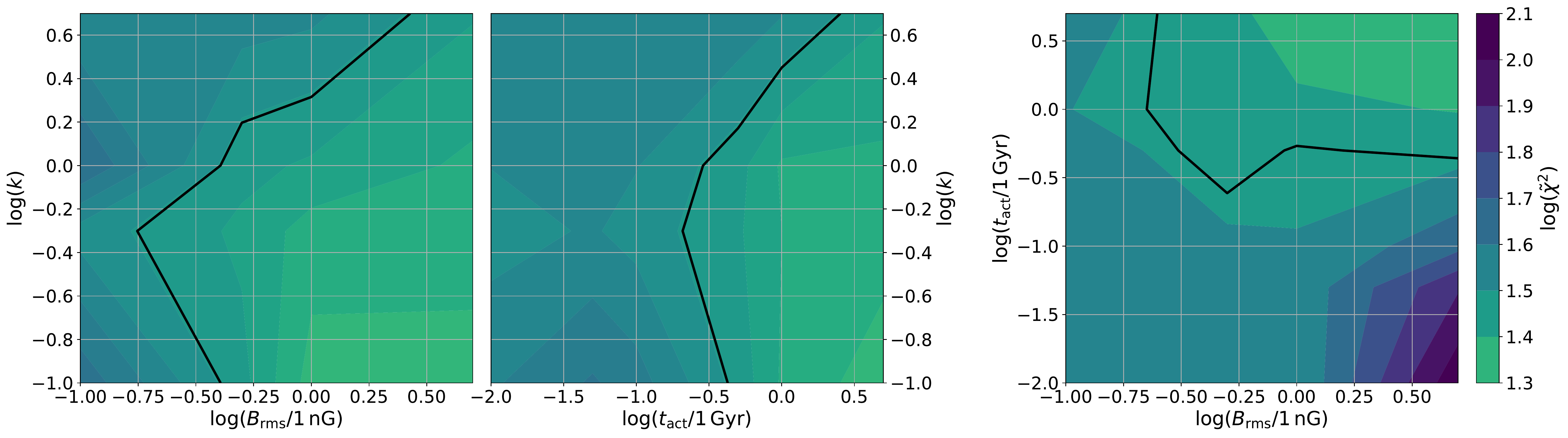} 
\caption{The marginalised (over the third parameter) chi-squared distribution that results from the optimization routine (\emph{upper panel}) as well as after including the residuals by the quadrupole data (\emph{lower panel}). Here the local source sample S11 is used. The black lines correspond to a chi-squared value of $20$ (dashed) and $30$ (solid), respectively.}
\label{fig:distrChi2}
\end{figure}

Using the source sample S11 consisting of eleven sources as well as the diffuse
contribution from low- and high-luminosity radio galaxies, the upper panel
of Fig.~\ref{fig:distrChi2} shows that the parameter space of
$(k,\,t_{\rm act},\,B_{\rm rms})$ is hardly constrained by the data, with the
exception of preferring a rather small leptonic energy density ($k<1$)
of these sources.  Since it is not possible to single out a best-fit
in the $(k,\,t_{\rm act},\,B_{\rm rms})$ space, we account in addition for the
quadrupole anisotropy in the post analysis: Keeping the fixed parameters
given by the optimization routine, we add the residuals of the quadrupole data,
obtaining an enhanced chi-squared value $\tilde{\chi}^2$. Here we only account for the averaged quadrupole amplitude $Q=\sqrt{\sum_{ij}Q_{ij}^2/9}$. Note that the observed quadrupolar components are currently not significant in any of the considered energy bins, so that the actual quadrupole strength might also be significantly smaller than what is used in the following. 
As shown in the lower panel of Fig.~\ref{fig:distrChi2}, the inclusion of the averaged quadrupole amplitude yields a 
preference for long lifetimes of low-luminosity sources as well as a large
EGMF strengths, emphasizing the importance of the quadrupole data in case of
a larger local source sample.  Strong EGMF fields combined with
short source lifetimes are strongly disfavoured, with $\chi^2\gtrsim 100$. 

\begin{figure}[htbp]
\centering
\includegraphics[width=.95\linewidth]{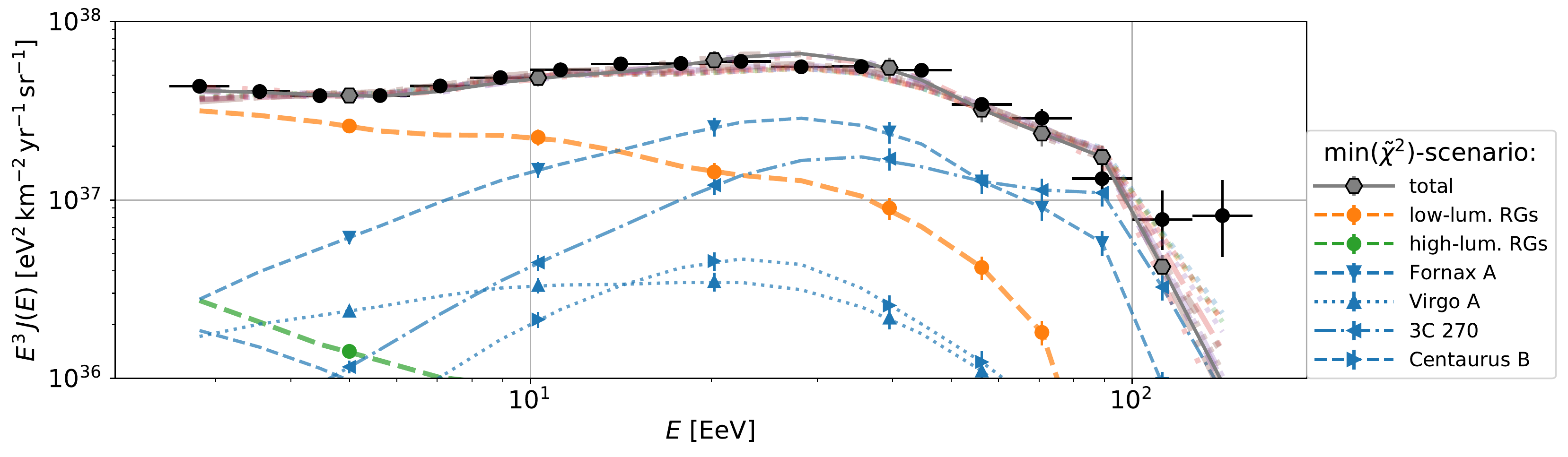}
\includegraphics[width=.575\linewidth]{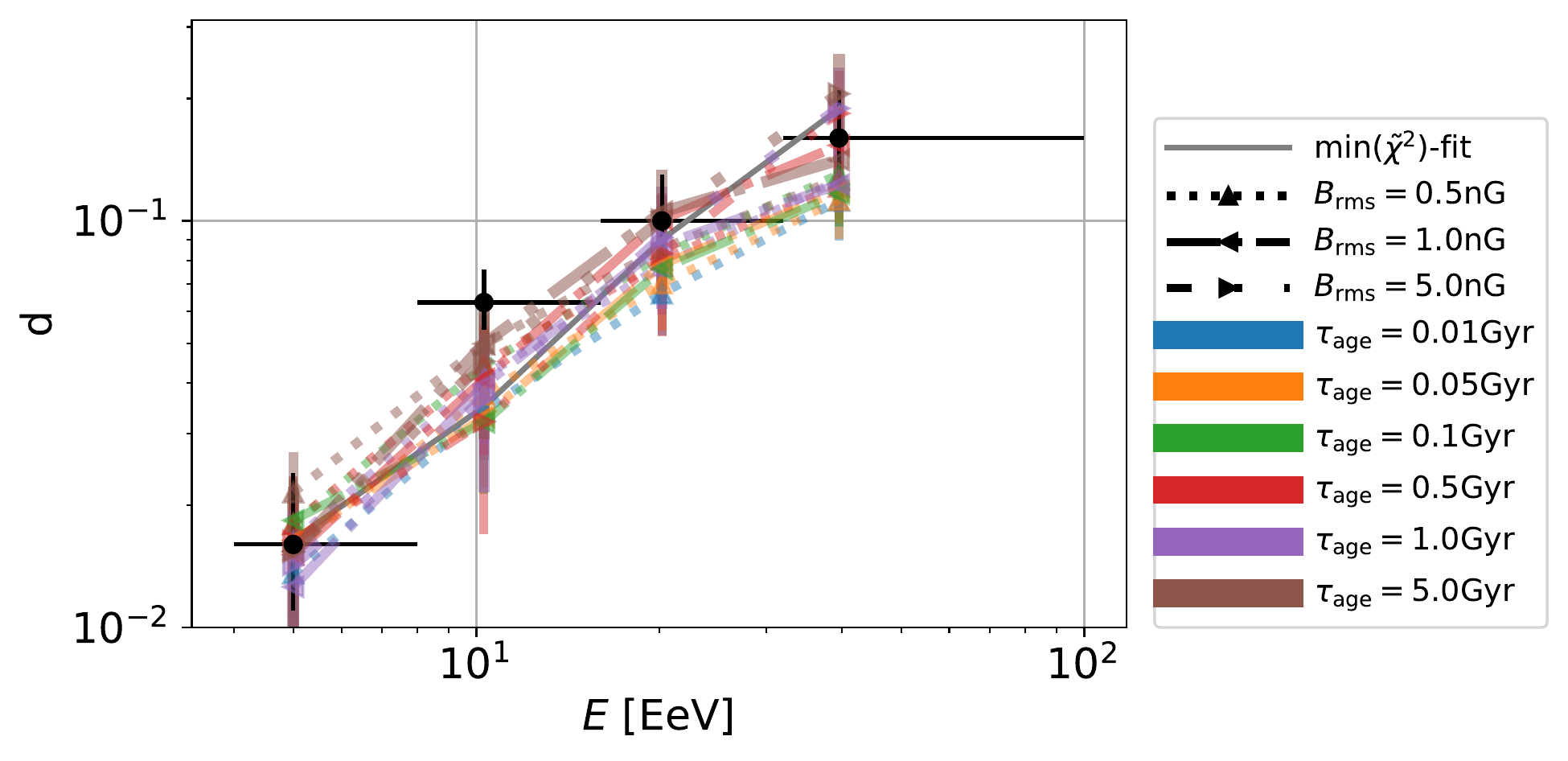} 
\includegraphics[width=.412\linewidth]{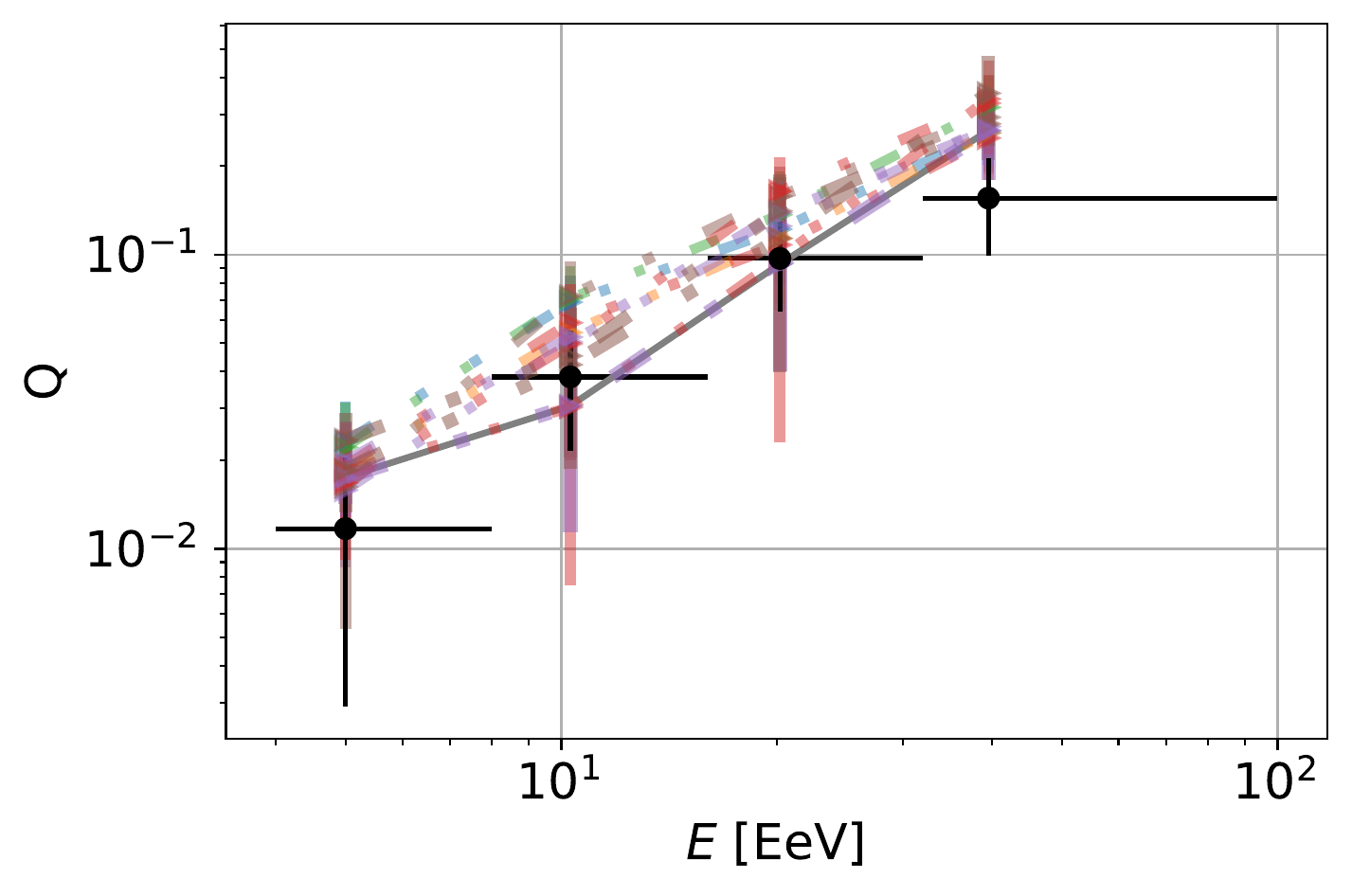}
\includegraphics[width=.489\linewidth]{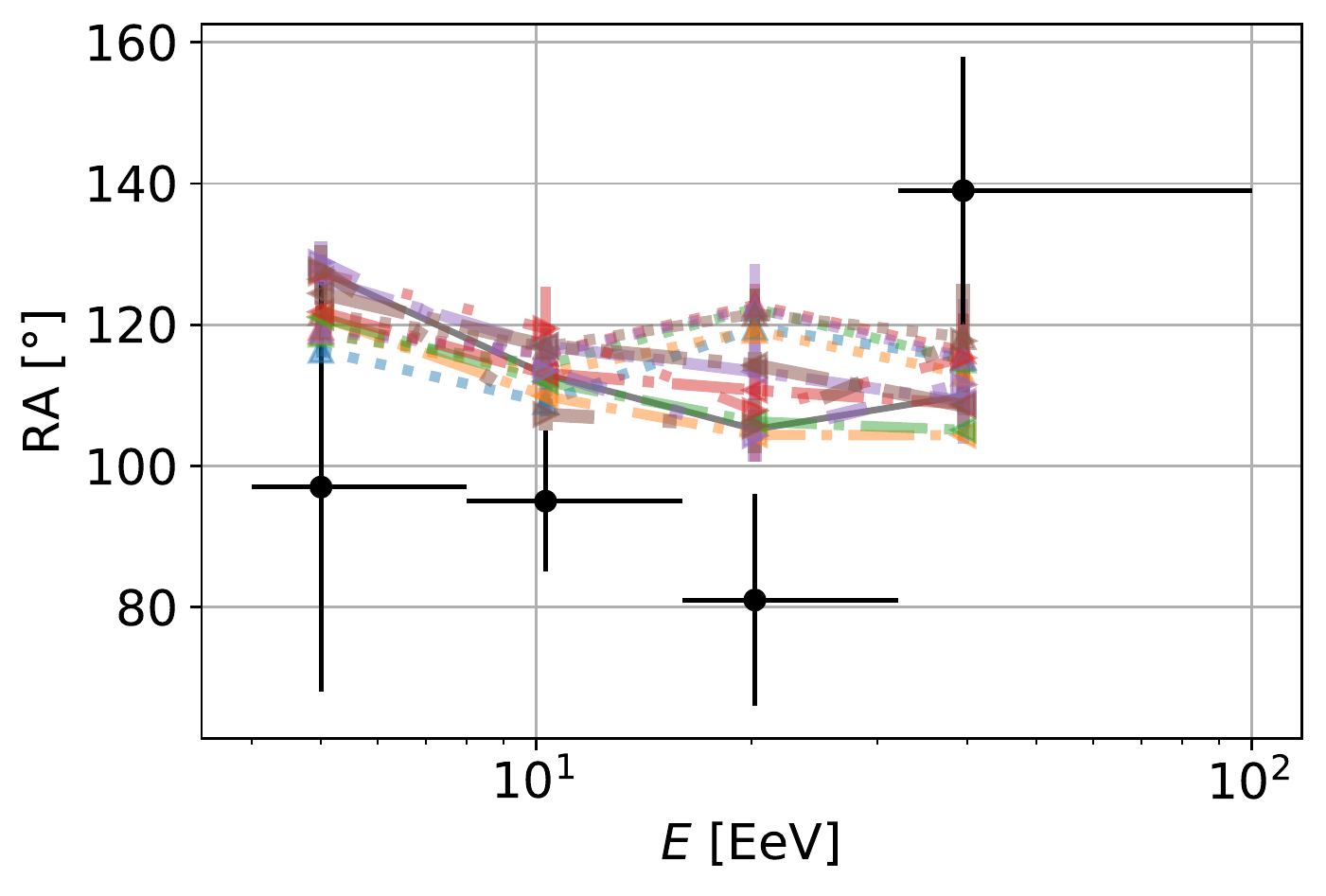}
\includegraphics[width=.489\linewidth]{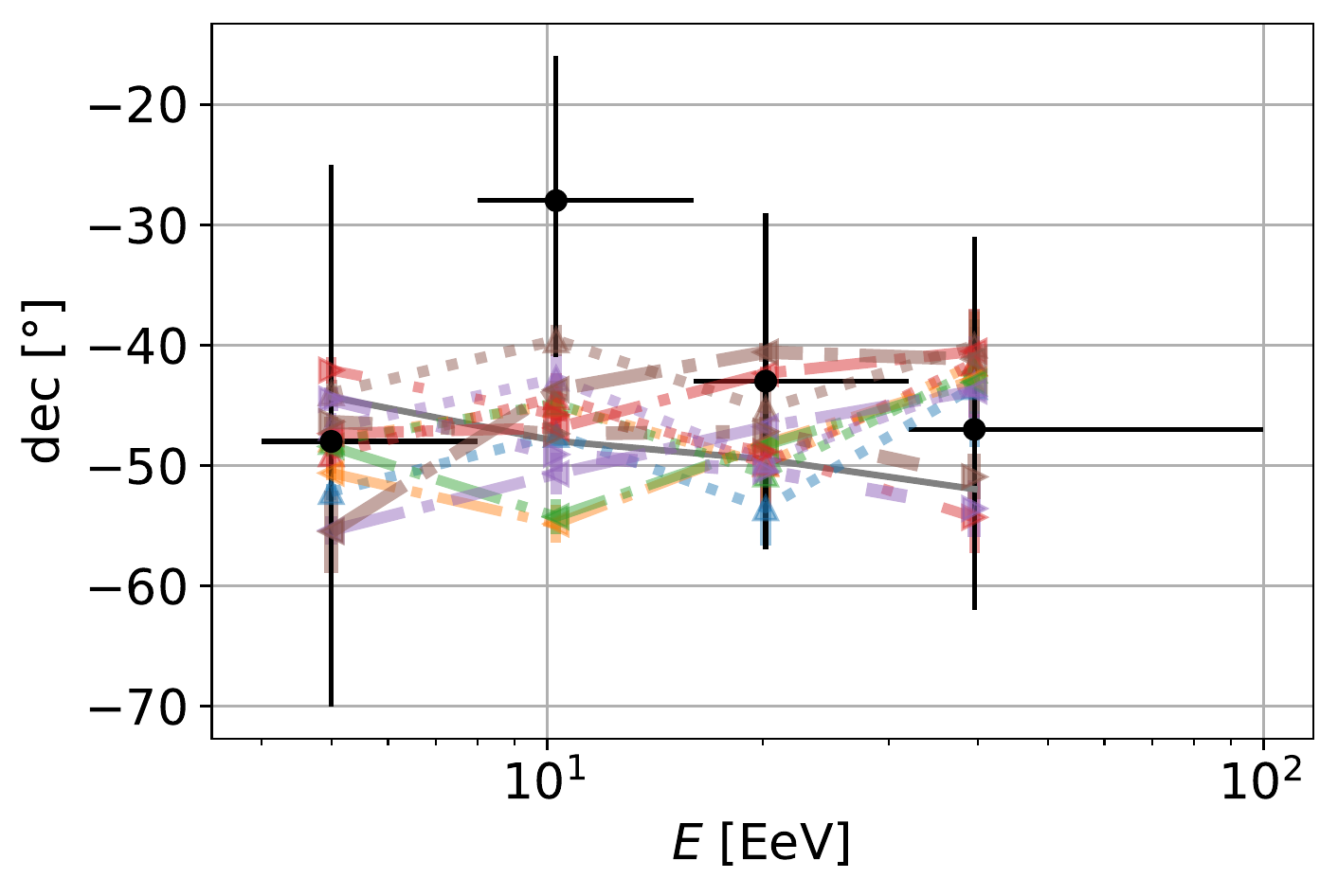}
\caption{The best-fit model predictions using the S11 local source sample and the corresponding observational data (black markers) of the energy spectrum \cite{PierreAuger:2020qqz} and the anisotropy characteristics \cite{deAlmeida+2021_ICRC}, respectively. 
\emph{Upper panel:} The resulting energy spectrum for the scenario with the smallest $\tilde{\chi}^2$ value including its different contributions as well as the total spectra for all scenarios yielding $\tilde{\chi}^2<30$ (dimmed colored lines without markers). The markers of the model prediction indicate those energies that have been taken into account in the in the chi-squared calculation.
\emph{Middle panel:} The resulting dipole (\emph{left}) and quadrupole (\emph{right}) amplitude for all scenarios with $\tilde{\chi}^2<30$ (\emph{right}). \emph{Lower panel:} The corresponding dipole direction in right ascension (\emph{left}) and declination (\emph{right}) of the different scenarios.}
\label{fig:fitsS11_wQ}
\end{figure}

Even including the quadrupole data,
there is a multitude of scenarios allowed. Investigating the relative
contribution of the different sources requiring an acceptable agreement
with the data (selected by $\tilde{\chi}^2<30$)\footnote{This choice is guided by the total number of data points but still a rather arbitrary one. Since the degrees of freedom of the chi-squared distribution are hard to estimate as the fit parameters are not independent of each other and the effective number of relevant parameters can diverge significantly from the total number of parameters, so that the actual confidence interval is unclear. However, varying the chosen chi-squared limit by $\Delta\tilde{\chi}^2\sim \pm 5$ does not change our conclusions at all.}, we recognize the following (see Fig.~\ref{fig:SrcCont_wQ}):
\begin{figure}[htbp]
\centering
\includegraphics[width=.95\linewidth]{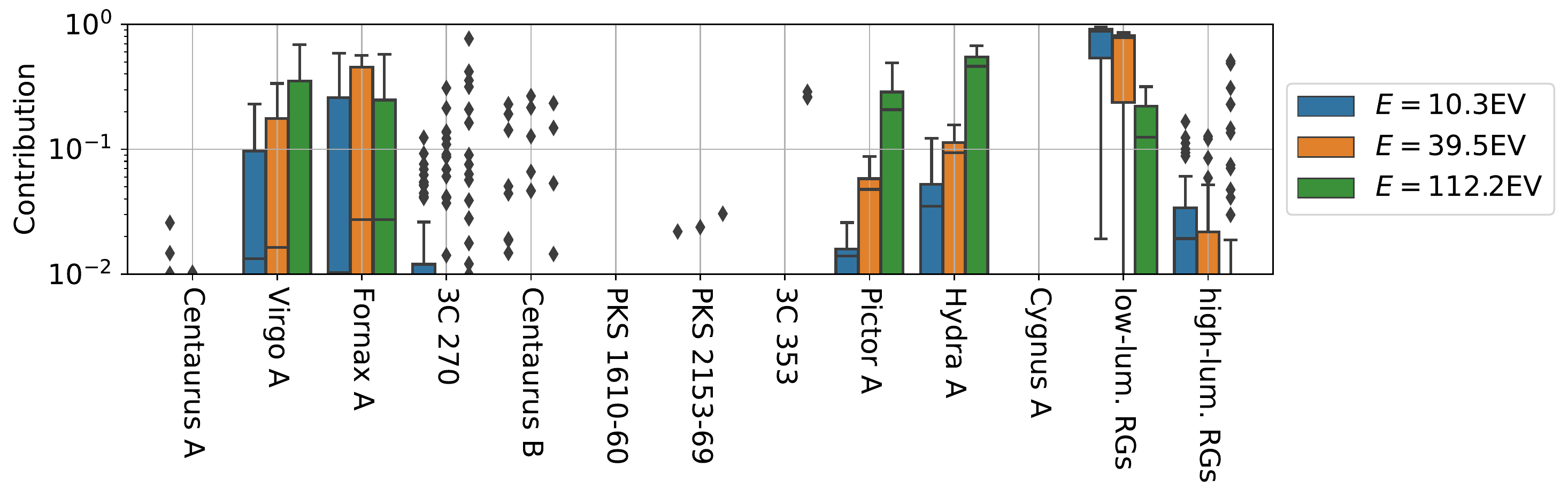} 
\caption{The relative contribution of the different sources to the UHECR flux at three different energies. Only results with $\tilde{\chi}^2<30$ are included. The solid line within the boxes refers to the median, the box represents the interquartile range (difference between the 75th and 25th percentiles of the data), the minimal and maximal values (excluding outliers) are indicated by the error bars, and outliers are shown as diamonds.}
\label{fig:SrcCont_wQ}
\end{figure}
\begin{figure}[htbp]
\includegraphics[width=.75\linewidth, trim=4cm 0 0 0]{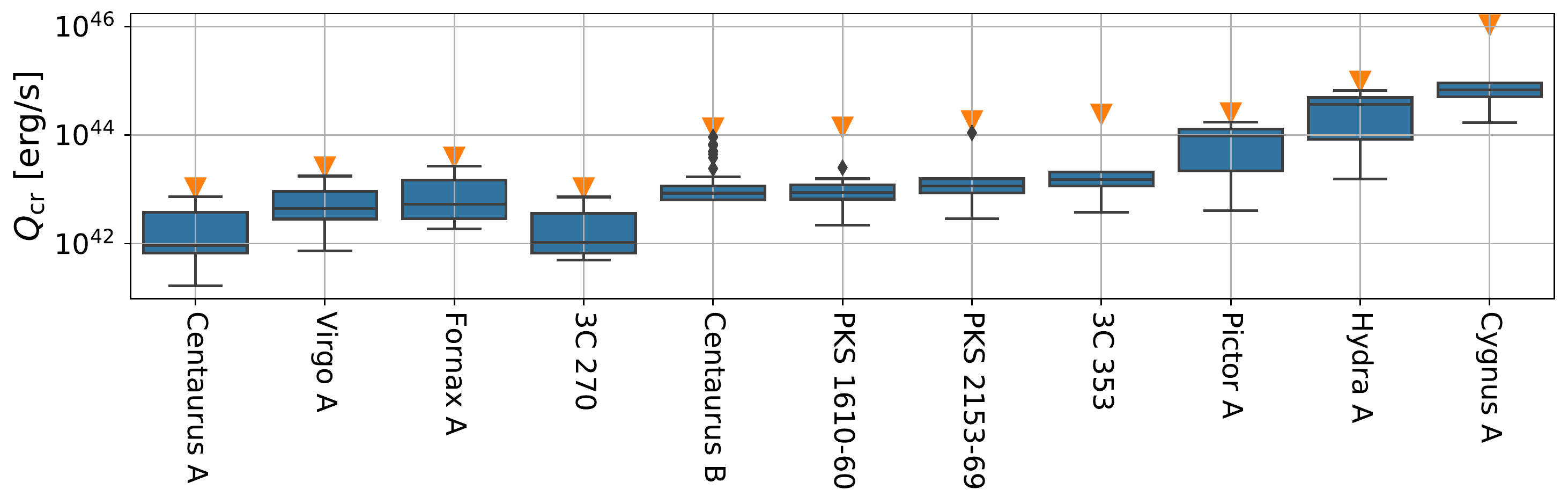}
\centering
\includegraphics[width=.87\linewidth]{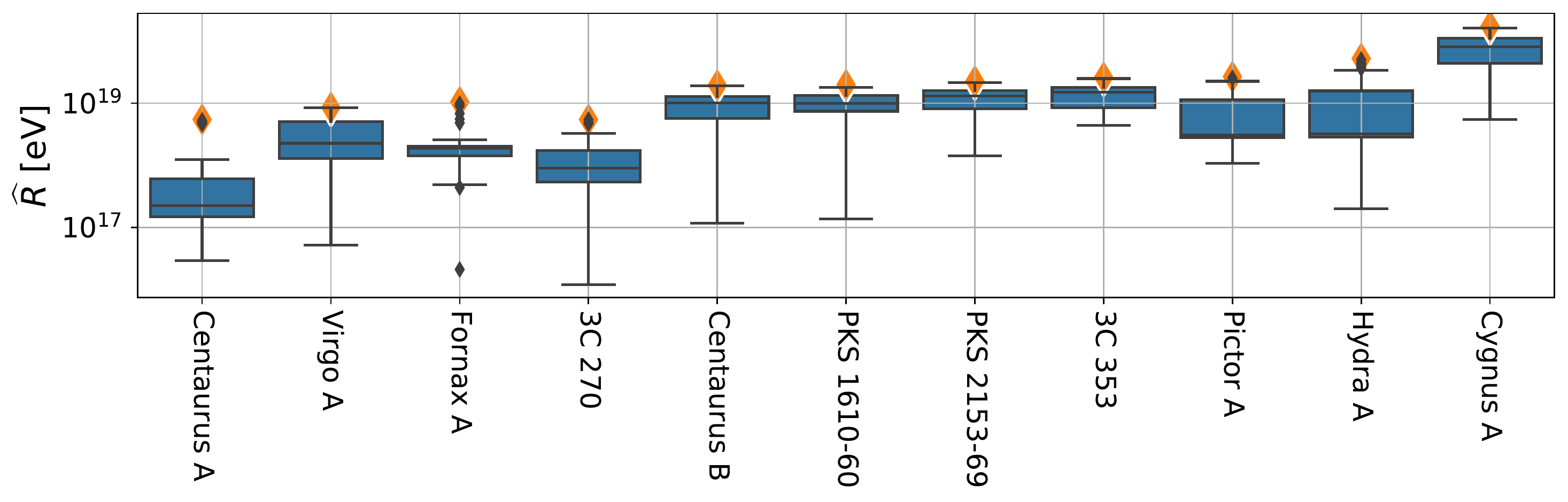}
\includegraphics[width=.115\linewidth, trim=0 -3.9cm 0 0]{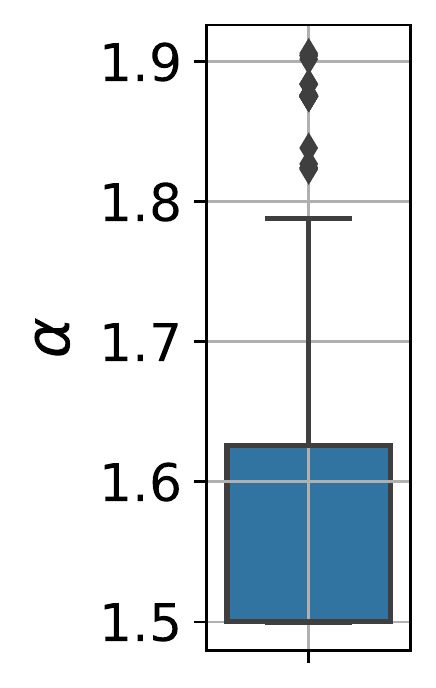}
\caption{The distribution of the CR power (\emph{upper panel}) and the maximal rigidity (\emph{lower left panel}) of individual radio galaxies of the local source sample S11 as well as the distribution of the source spectral index (\emph{lower right panel}). Only results with $\tilde{\chi}^2<30$ are included. The orange triangle in the upper panel indicates the underlying jet power $Q_{\rm jet}$, and the orange diamond in the lower panel gives the parameter independent rigidity $\sqrt{Q_{\rm jet}/c}$. Details on the boxplot features are given in the caption of Fig.~\ref{fig:SrcCont_wQ}.}
\label{fig:SrcFeatures_wQ}
\end{figure}
On average, the diffuse contribution of low-luminosity sources dominates at
most energies, in particular at $E\ll 100\,\text{EeV}$,  while the majority
of the local sources in the S11 sample contribute less than $1\,\%$.
Surprisingly, also Centaurus~A contributes only a negligible fraction to the
total UHECR intensity, especially at high energies, while the most significant
local sources on average are Virgo~A, Fornax~A, Pictor~A, and Hydra~A.
The latter two are important in particular at the highest energies,
$E\sim 100\,\text{EeV}$. The relative dominance between those four sources
depends significantly on the EGMF strength and their lifetimes. There are
however other scenarios where sources such as 3C\,270, Centaurus~B or 3C\,353 contribute a significant
percentage of the UHECR flux, as visible in the Fig.~\ref{fig:SrcCont_wQ} or
the upper panel of Fig.~\ref{fig:fitsS11_wQ}.
The best-fit scenarios also exclude a major contribution of several local
sources, such as Centaurus~A (especially at high energies) and Cygnus~A
(especially at low energies)---what mostly results from their extreme
distances to Earth (as well as the relative short lifetime in case of Cygnus~A)
as further elaborated in Sect.~\ref{sec:disc}. 
Figure~\ref{fig:SrcFeatures_wQ} displays the mean and spread of the 
main source characteristics (CR power, maximal rigidity, spectral index) of
the fits with $\tilde\chi^2<30$ for the S11 sample. On average, their CR
power is about an order of magnitude smaller than the given jet power,
and the maximal rigidity is mostly smaller by factor of about two to five
than $\sqrt{Q_{\rm jet}/c}$, An exception is Centaurus~A, whose median
maximal rigidity is the smallest of all local sources. Note that the
reference value of the individual jet power depends for most sources
(exceptions are Centaurus~A, Cygnus~A, Pictor~A and 3C\,353) on the supposed
radio-jet-power correlation~\cite{Ineson+2017}. Further, we obtain a clear
indication of a hard source spectrum of $\alpha\sim 1.5$, which has already
been the case for the S5 results. The preference for rather hard spectra
is in line with previous results which used an ``average source''
approach~\cite{PierreAuger:2016use,Hooper:2009fd}. Allowing for an even harder source spectrum the optimization procedure will also deliver a harder source spectrum on average, however, the corresponding chi-squared-values improve only marginally. Hence, our fit results are not very sensitive to the choice of $\alpha$, and we could fix $\alpha$ at some value in the range $1\lesssim \alpha \leq 2$ without introducing any major differences in the outcome. 
Looking at the corresponding fits to the data in Fig.~\ref{fig:fitsS11_wQ},
we recognize that our model yields for almost all (accepted) scenarios a
fairly accurate description of the intensity spectrum, even at energies
below the ankle which are outside our fit range. In addition,  the energy
dependence
of the dipole and quadrupole strength is reproduced well, where one
should keep in mind again that the latter has not even been accounted for
by the optimization routine. The clearest disagreement emerges from
the dipole direction as our local source sample can hardly account for
those strong non-monotonous changes, such as for the right ascension
above $32\,\text{EeV}$ or the declination between 8 and 16\,EeV, that
are indicated by the data. Further, there is a systematic difference of
about 20$\degree$ in right ascension between the observed dipole and our
predictions. However, in most cases we are still within or at most a few
degrees off the uncertainty range of the observations (for a more detailed
discussion on this issue see Sect.~\ref{sec:disc}). 
\begin{figure}[htbp]
\centering
\includegraphics[width=.49\linewidth]{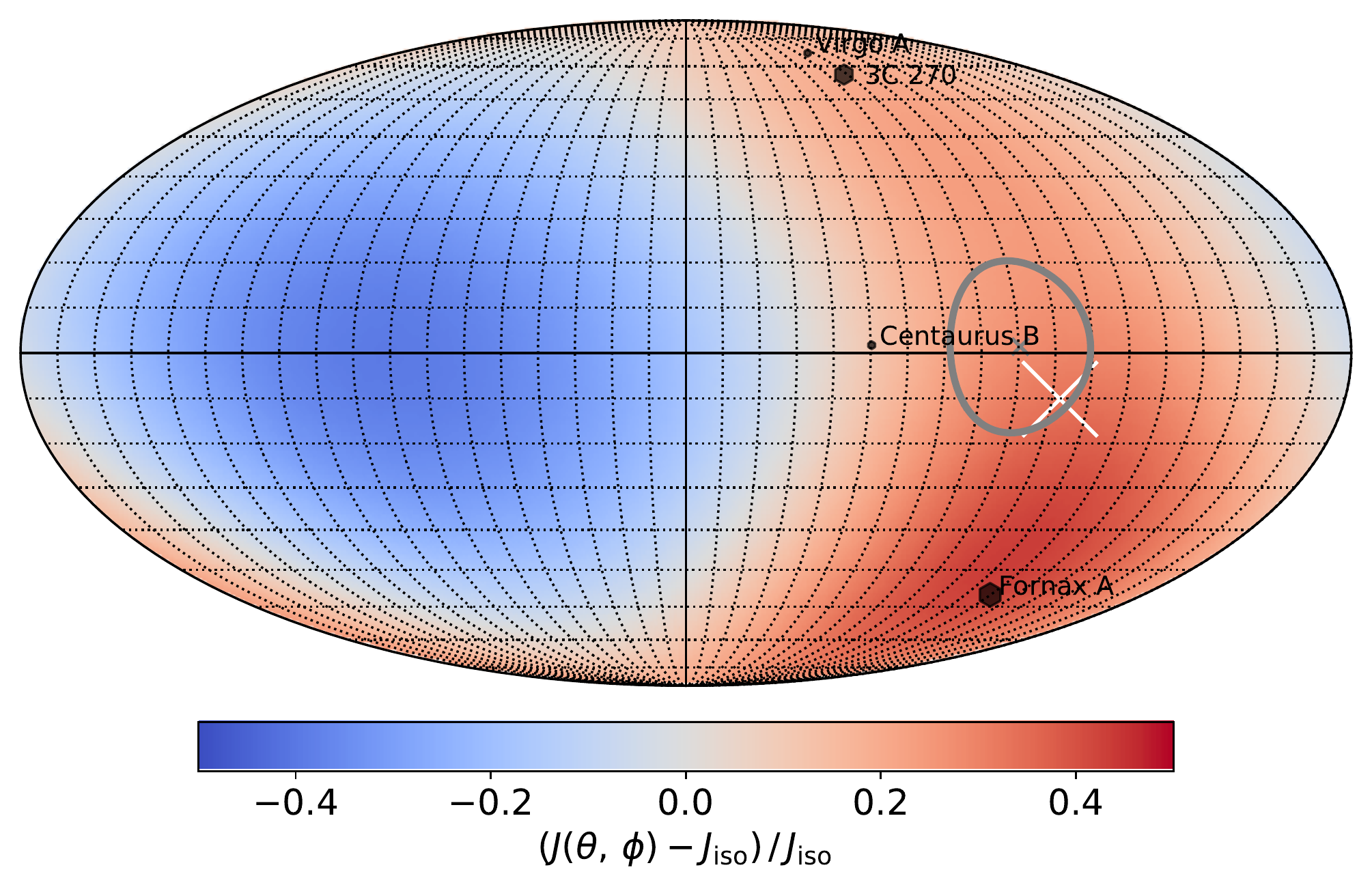} 
\includegraphics[width=.49\linewidth]{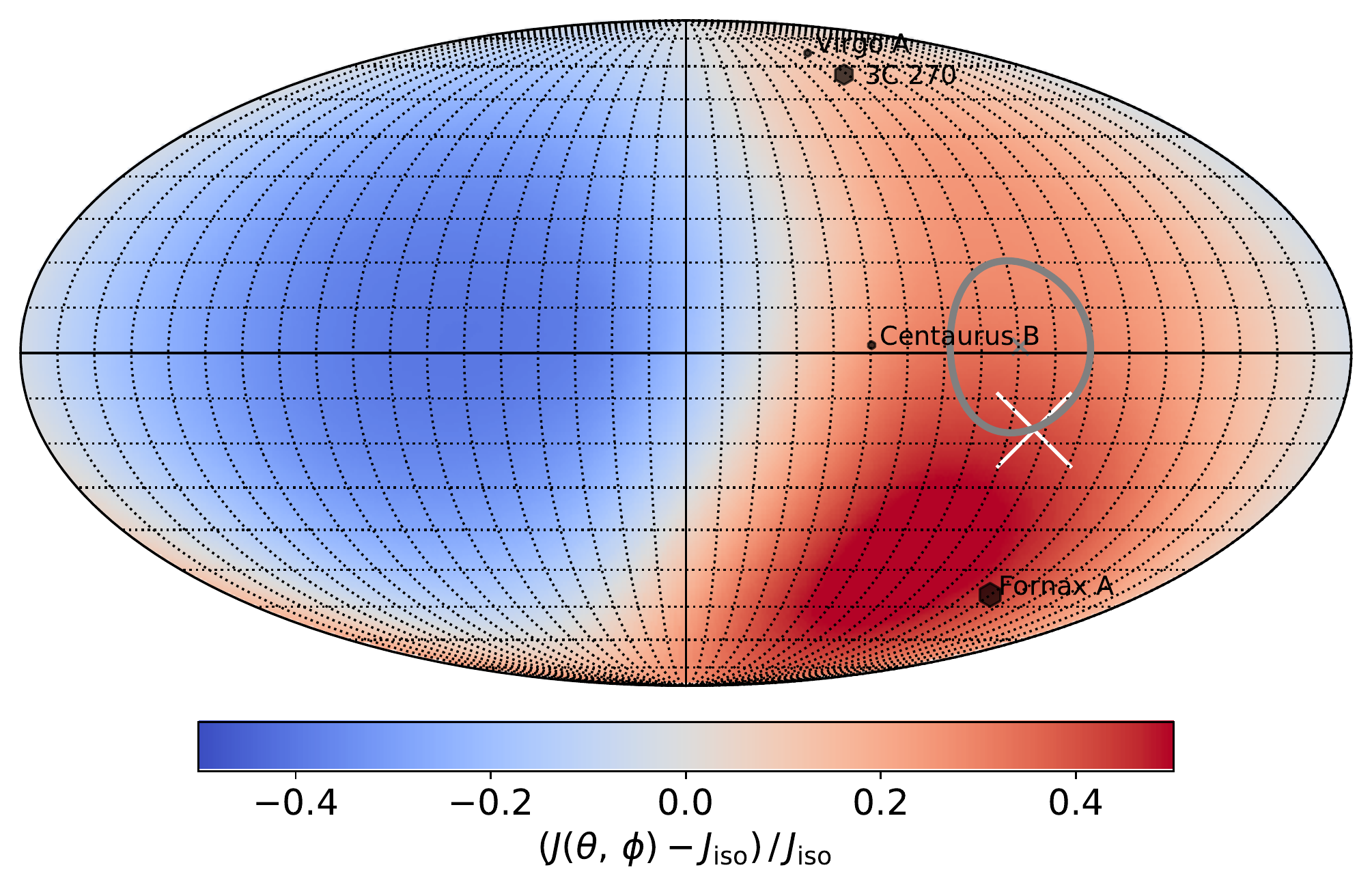} 
\caption{Mollweide projection in Galactic coordinates of the UHECR arrival directions at 40\,\text{EeV} averaged on top-hat windows of $45\degree$ radius. Shown is the min($\tilde{\chi}^2$)-scenario based on the S11 local source sample before (\emph{left}) and after (\emph{right}) propagation through the Galactic magnetic field. The white cross shows the resulting dipole direction, whereas the observed dipole direction and its uncertainty at about this energy ($>32\,\text{EeV}$) is indicated by the grey cross and the surrounding solid line. Sources that provide more than $10\%$ of the flux at this energy are indicated by black hexagons (their size scales with the source contribution). 
\label{fig:BFskymap40}}
\end{figure}
We can conclude that a local source sample consisting of either Hydra~A
and Pictor~A or Virgo~A/ 3C 270 and Fornax~A is sufficient to explain the data.
Hence, the local source sample can in principle be reduced to a combination
of these two sources without a significant change in the goodness of the fit.
But, it needs to be emphasized that such a scenario requires an EGMF strength at the order of nG (for 1\,Mpc coherence length)\footnote{Note that Virgo~A
  is located close to the Supergalactic plane, where the magnetic field
  strength can be at the order of a few tens of nG.} as well as a
long lifetime of about Gyrs or more. Analysing the angular distribution of the arriving CRs at the highest energies (see Fig.~\ref{fig:BFskymap40}) it becomes clear that these fit scenarios lead to a flux deficit at about the whole hemisphere at low Galactic latitudes ($l\lesssim 180\degree$). Further, we illustrate the impact of the Galactic magnetic field, which shifts the dipole direction by about $10\,\degree$ in case of this particular scenario.
Next, we will test if a substantial increase of the local source sample of
up to 26 sources (hereafter S26), including several sources at a Galactic
longitude $l\sim150\degree$ close to the Galactic plane, introduces some
substantial changes to the previous outcome.

\subsubsection{Twenty-six local sources}

\begin{figure}[htbp]
\centering
\includegraphics[width=.489\linewidth]{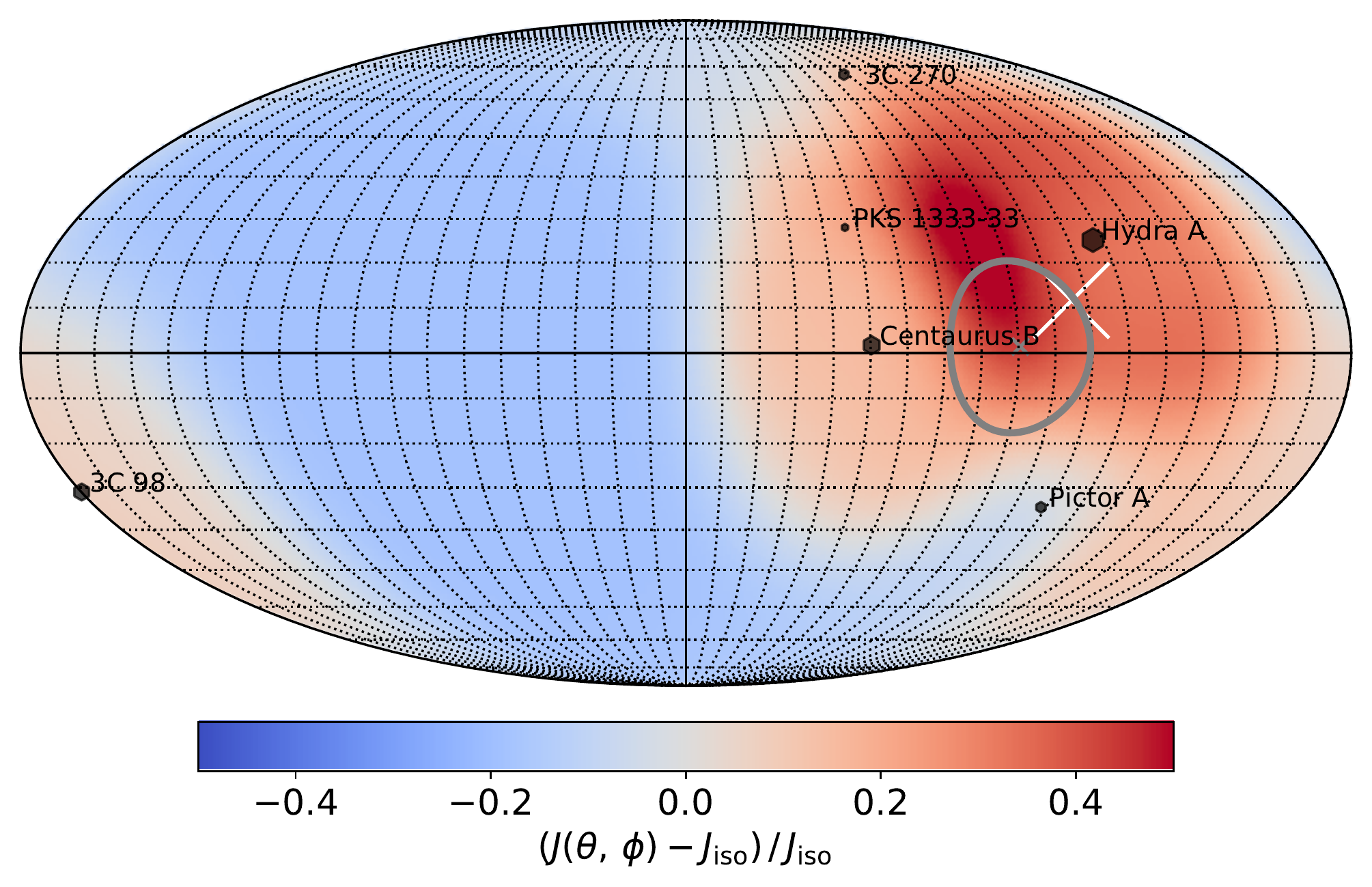}
\includegraphics[width=.489\linewidth]{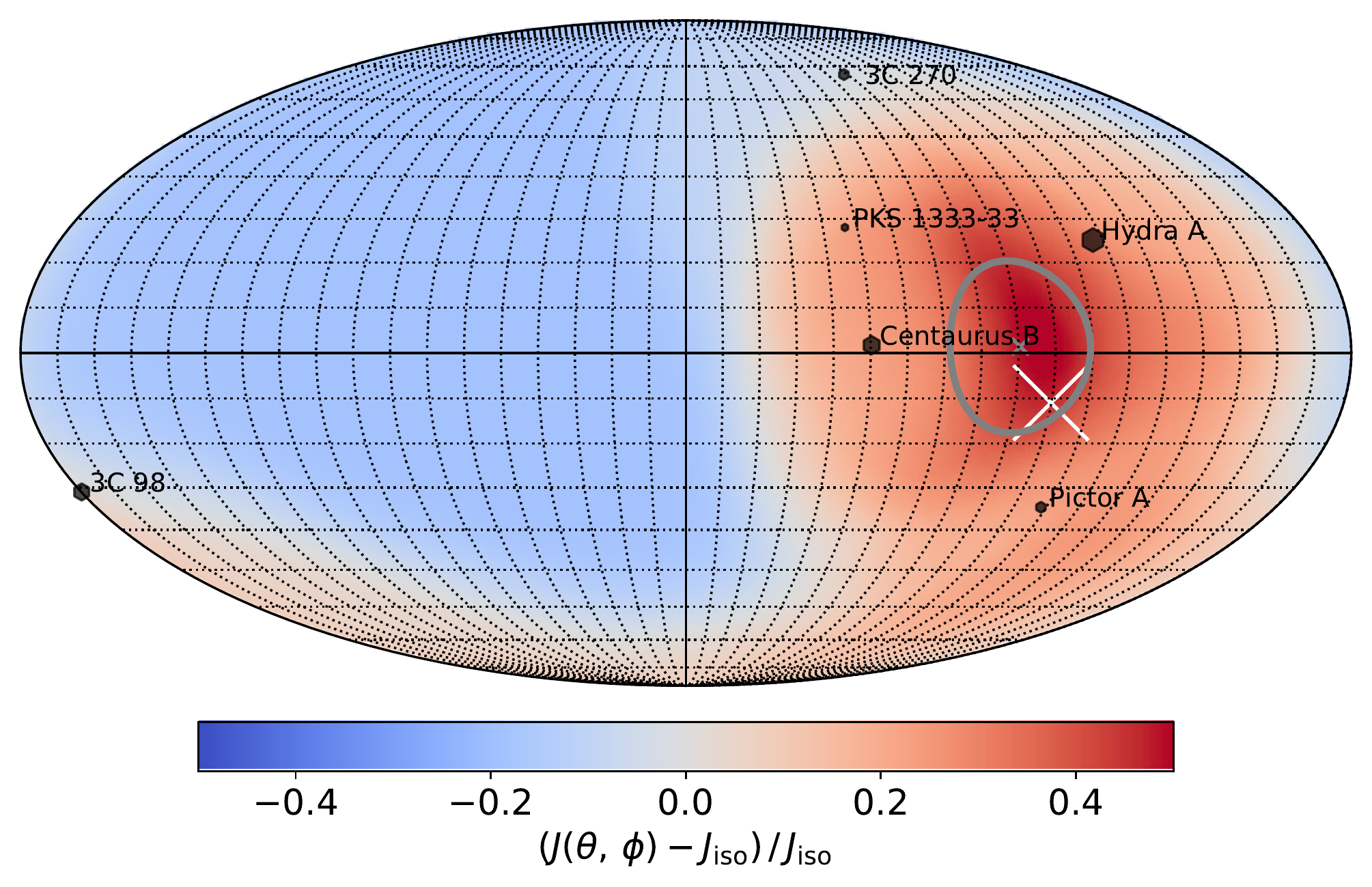} 
\caption{Mollweide projection of UHECR arrival directions at 40\,\text{EeV} in Galactic coordinates averaged on top-hat windows of $45\degree$ radius. Shown is the best-fit scenario based on the S26 local source sample for $B_{\rm rms}\simeq 1\,\text{nG}$, $t_{\rm act}\simeq 10\,\text{Myr}$ and $k=1$ before (\emph{left}) and after (\emph{right}) propagation through the Galactic magnetic field. Markers and lines refer to the same properties as given in Fig.~\ref{fig:BFskymap40}.}
\label{fig:BFskymap40_S26}
\end{figure}

\begin{figure}[htbp]
\centering
\includegraphics[width=.565\linewidth]{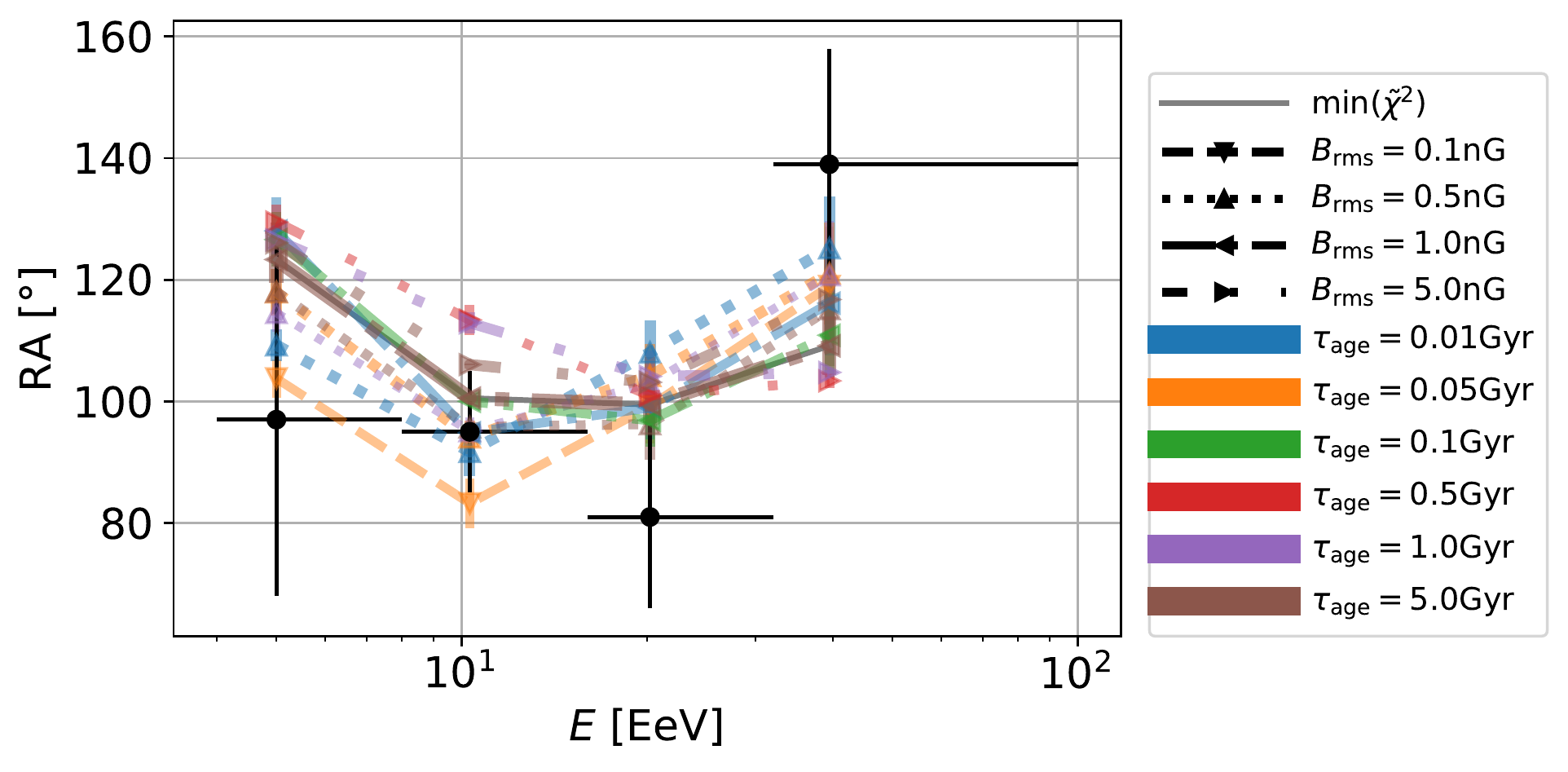}
\includegraphics[width=.415\linewidth]{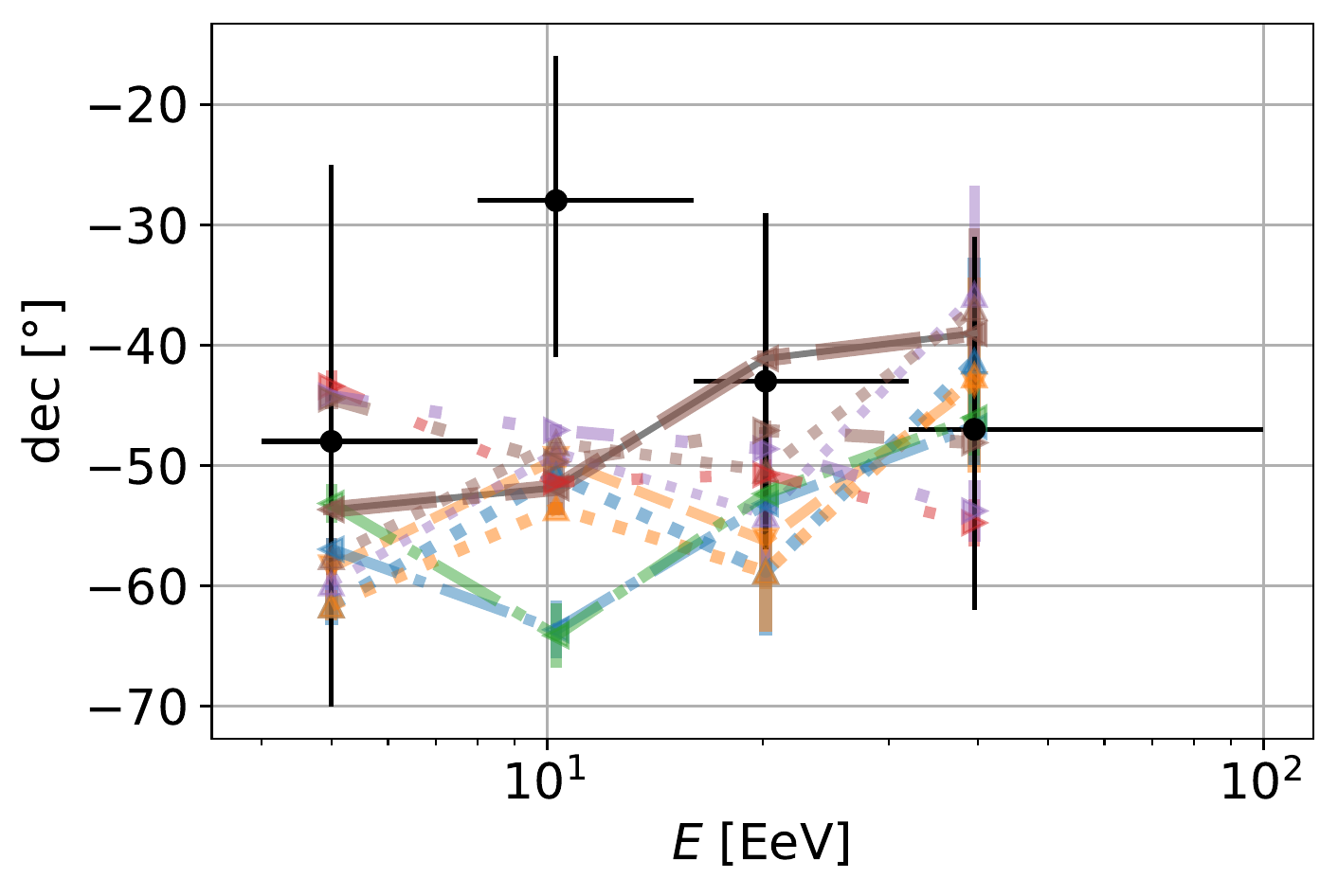}
\caption{The best-fit results of the dipole direction in right ascension (\emph{left}) and declination (\emph{right}) using the S26 local source sample.}
\label{fig:fitsS26_wQ}
\end{figure}

Although the goodness of the fit does not improve significantly by doubling
the size of the local source sample, the use of the S26 sample introduces
some major differences with respect to the previous results: There are several
fits with $\tilde{\chi}^2<30$ for a short source lifetime,
$t_{\rm act}\sim 10\,\text{Myr}$, and there is much more variety in the
contribution by the different sources: 
First, there are still several scenarios (for $B_{\rm rms}\sim 1\,\text{nG}$ and
$t_{\rm act}\sim 1\,\text{Gyr}$) that suggest that Fornax~A and Virgo~A are the
dominating sources. However, in addition a minor contribution on the order
of about 10\% at 40\,EeV by sources such as 3C\,98, PKS 1333-33, 3C\,129 or 3C\,111 is introduced. 
Second, we obtain multiple scenarios, where there is no need for a contribution
by Fornax A and Virgo~A. 
In these cases, typically some combination of the
local sources Hydra~A, Centaurus~B, 3C\,98, 3C\,270, Pictor~A and
PKS 1333-33 can explain the data. Figure.~\ref{fig:BFskymap40_S26} gives an
illustrative examples of the CR arrival directions at 40\,EeV for such a
scenario with and without the impact of Galactic magnetic field deflections.
In addition, the use of the S26 source sample leads to some improvement of the resulting anisotropy features, in particular the right ascension of the dipole direction due to the contributions of 3C\,129 and/or 3C\,98, as can be seen in Fig.~\ref{fig:fitsS26_wQ}. In general, the increase of the local source sample allows for a stronger energy dependence of the dipole direction, but the
agreement with the observed declination of the dipole direction between 8 and 16\,EeV is still not perfect. 

\section{Discussion}
\label{sec:disc}

Based on the individual properties of local radio galaxies, we have developed
a model for their possible UHECR contribution at Earth. In the following,
we discuss first the main assumptions we have used:
\begin{enumerate}
\item[(i)] Magnetic field: The most significant influence on our outcome
  results from the impact of the EGMF and the GMF on the CR propagation. The
  chosen uniform and isotropic turbulent EGMF does not introduce any
  systematic shift of the UHECR arrival directions, but only spreads the
  angular distribution. We have adopted this simplified approach since EGMF
  models based on magnetohydrodynamics simulations differ strongly in their
  predictions~\cite{Hackstein+2018}. Furthermore, the structure of a possible
  regular component which may shift the UHECR arrival directions in
  particular from sources in the Supergalactic plane, cannot be predicted
  by such simulations. In addition, the effective field strength of the
  EGMF will depend on the source location, such that CRs from sources
  within the Supergalactic plane have to cope with a significantly
  higher field strength than those that originate outside the plane. However, using a non-uniform EGMF model also leads to the ambiguity of the exact source location within this magnetic field structure causing an additional source of uncertainty. 
  Including such a non-uniform, turbulent field could change the dipole direction and lead to some additional
  contribution, e.g.\ by rather distant sources outside the Supergalactic plane, such
  as Cygnus~A or 3C\,353, that feature short lifetimes. 
  But in general, the current (and most likely future) lack of information on the EGMF will always yield the major source of uncertainty with respect to any kind of UHECR source prediction.
  \\
  In contrast to the used EGMF, the GMF has a significant effect on the dipole
  direction due to its regular field component. While the used JF12 model is
  the most sophisticated attempt to model the GMF, it is built on a limited set
  of information and has several weaknesses~\cite{Kleinmann+2019,Beck+2016}.
  Therefore  the shift of the dipole direction predicted in the JF12 model
  has a rather large systematic error which we have not included in our fit
  procedure.
\item[(ii)] Mean modification factor: The impact of CR interactions
  with photons from the CMB and the EBL are incorporated by calculating
  the ratio of the modified and the unmodified spectral rigidity distribution
  for a given set of parameters. The modified distribution is obtained from
  1D simulations, hence without including the increased travel time of CRs
  caused by magnetic field deflections. Further, we only account for the
  elemental mean of this factor to become independent of the unknown source
  abundances of different CR elements. This approach yields a significant
  reduction of the parameter space we have to evaluate but introduces
  uncertainties up to about a factor of five for small source distances and
  rigidities (cf.\ with the right panel of Fig.~\ref{fig:interactionEffects}).
  Thus, the closer the individual radio galaxy, the higher is the chance of
  an additional modification of its flux contribution at rigidities
  $<10\,\text{EV}$, whereas at the highest rigidities the mean modification
  factor $\bar{\eta}$ is fairly accurate. 
\item[(iii)] Elemental composition of UHECRs at Earth: Using the mean
  modification factor allowed us to determine the rigidity dependent UHECR
  intensity $J_R$ at Earth independent of the initial source abundances.
  To compare our prediction with the observations, we have converted $J_R$
  into the energy space by using the observed composition data
  for $\langle \ln A \rangle_{\rm obs}$. For the considered energies above
  4\,EeV, where protons are mostly absent, it is appropriate to use
  $Z\simeq A/2$. Since our model predictions only depends on the CR rigidity,
  there is no need to include the information on the variance of the observed
  composition as we still have the freedom to choose, e.g., different CR mass
  numbers from different sources at the same energy, keeping
  $\langle \ln A \rangle$ fixes. 
\item[(iv)] Source model: We have determined the jet power and lifetime of
  individual high-luminosity radio galaxies from a recent dynamical
  evolution model~\cite{Machalski+2021}, whereas the jet power of the other
  sources (except for Centaurus~A) has been estimated from the well-established
  correlation between radio and jet-power~\cite{Ineson+2017}. Although this
  correlation holds predominantly for FR-II sources, observations indicate
  a similar correlation for low-luminosity sources~\cite{GodfreyShabala2013}.
  The rather large uncertainties and deviations from the given correlation 
  for individual low-luminosity radio galaxies is partly compensated
  by allowing for an individual value of the fraction $g_{\rm m}$ of jet
  and matter power.  Similarly, this parameter can on average also account
  for a possible enhanced activity of the individual sources in the past, such
  as indicated for Centaurus~A~\cite{Wykes+2013, Eilek2014} and
  Fornax~A~\cite{Lanz:2010ai}. 
\end{enumerate}
Based on this model, we are able to predict the anisotropic UHECR intensity
as a sum of contributions from individual local radio sources plus the
isotropic contribution from low- and high-luminosity radio galaxies.
A good agreement with the observational data---including
the energy spectrum below the ankle as well as the quadrupole strength which
where both not included in the fitting procedure---is already obtained by
using only Fornax~A as well as Virgo~A and/or 3C\,270 as
local sources. However, in this scenario the EGMF strength has to be around
$B_{\rm rms}\sim 1\,\text{nG}$ (for a coherence length of 1\,Mpc), while the
lifetimes of the sources have to be very large, of the order of Gyrs.
In these cases, the ankle of the CR spectrum often marks the transition from
a dominant contribution by the bulk of low-luminosity radio galaxy
(at $E\lesssim 5\,\text{EeV}$) towards the dominance by local sources
(at $E\gtrsim 5\,\text{EeV}$). The diffuse contribution by high-luminosity radio
sources is on average at a level of a few percent at most.

Including more local sources---in particular at Galactic longitudes between
90\degree\ and 180\degree---leads to a multitude of possible scenarios where,
e.g., Hydra~A, Pictor A or Centaurus~B can become the dominating UHECR sources
without improving the fit significantly. Hence, data on the large-scale
anisotropy of UHECRs alone do not allow one to single out a specific set
of dominant local source of CRs. However, our analysis shows that
at least one of the previously mentioned sources will have a major impact
on the UHECR data, if radio galaxies are the sources of UHECRs. We could also
exclude a \emph{major\/} UHECR contribution from plenty of local sources:
Perhaps most surprisingly Centaurus~A and Cygnus~A are among these disfavoured
sources. The latter shows a quite short jet lifetime, leading to a strong
suppression at low energies due to the deflections by the EGMF. Centaurus~A
on the other hand is rather the opposite case, being very close and showing
a comparatively low CR power. Therefore a strong CR anisotropy is introduced
by this source even for the strongest EGMF cases which can hardly be
compensated by other local sources without generating a strong quadrupole
anisotropy.

Our predicitions, as those of all other models aiming to explain the UHECR
compositional data, depend on the hadronic  interaction model applied to
interpret the $X_{\rm max}$ data on the depth of the shower maximum. We
decided to present the results in Sect.~\ref{sec:FitResults} for the
interaction model EPOS-LHC, which yields in general the best agreements to
the data---with a minimal chi-squared values of $\text{min}(\chi^2)\sim 10$,
whereas the other interaction models typically lead to scenarios with
$\text{min}(\chi^2)\sim 20$. The main differences in our results caused
by using  Sibyll2.3c or  QGSJetII-04 can be summarized as follows:

Using the heavier chemical composition predicted by Sibyll2.3c yields quite
similar results as previously discussed except for the following major
differences: (i) Centaurus~A (and B) can contribute a significant fracion
of the observed UHECR intensity (especially at low energies), due to the
stronger magnetic field deflections of nuclei with higher charges; (ii)
the bulk of high-luminosity radio galaxies is on average more important
than the low-luminosity counterpart; and (iii) the source spectrum is on
average rather close to $\alpha\sim 2$. The major disagreement with respect
to the data is the right ascension of the dipole direction below 32\,EeV,
which shows deviations of up to $55\degree$ using S5. 

Using the lighter chemical composition prediction of QGSJetII-04 the
UHECRs suffer from significantly less deflections by the EGMF and the GMF.
Therefore the diffuse contribution by the bulk of low-luminosity radio galaxies
becomes the major UHECR sources at all energies, while the main local
sources (Fornax~A and Hydra~A) only contribute $10\%$ at most. Further, we
obtain the need for a small leptonic-to-hadronic energy ratio, i.e.\ $k<1$,
and a hard source spectrum. In this case, the major disagreement with respect
to the data is either a too weak dipole strength above 16\,EeV or a deviation
of the dipole's declination angle of up to $40\degree$ above 8\,EeV. In
addition, the inclusion of a larger number of local sources often leads to
too large a quadrupole anisotropy, although the fit to the other data
clearly benefits (leading to $\text{min}(\chi^2)\sim 10$) from using S26
instead of the smaller source samples.

Next, we want to compare our results with four indications for
medium-scale anisotropies at energies
$\geq 40\,\text{EeV}$~\cite{diMatteo+2021ICRC,diMatteo+2019ICRC}. Two of
these potential  excesses at about ($12^{\rm h}\,50^{\rm m}$, -50\degree) and
($1^{\rm h}\,50^{\rm m}$, -35\degree) in equatorial coordinates are shown
by black long-dashed and dash-dotted lines in Fig.~\ref{fig:srcSky},
respectively. We find for several scenarios, such as the one with the
smallest $\tilde{\chi}^2$ value, where Fornax~A as well as Virgo~A and/or 3C\,270 are the dominating local sources 
a quite good agreement with theses excesses. However, the significance of
these excess has declined~\cite{diMatteo+2021ICRC}.
Instead, another excess at about ($2^{\rm h}\,30^{\rm m}$, +50\degree) in
equatorial coordinates (black dotted line in Fig.~\ref{fig:srcSky}) has
emerged that requires sources close to the Galactic plane at a Galactic
longitude between $90\degree$ and $180\degree$.
Although the S26 sample features at least four low-luminosity sources
(3C\,129, 3C\,111, 3C\,83.1 and 3C\,66B) in this region---whereof 3C\,111
shows even some evidence of boosted emission~\cite{deJong:2012ty}---their
possible contribution to this excess depends strongly on the GMF model
and the rigidity of the particles at the highest energies. Using the JF12
model as well as a rather heavy chemical composition (according to the
hadronic interaction models EPOS-LHC or Sibyll2.3c) above 40\,EeV,
UHECRs from these sources are deflected by more than $30\degree$ towards
smaller Galactic latitudes, as can be seen for 3C\,98 in
Fig.~\ref{fig:BFskymap40_S26}.
In the directions of the fourth excess of events at about
($9^{\rm h}\,30^{\rm m}$, +54\degree) shown by the short-dashed line in
Fig.~\ref{fig:srcSky}, our sample contains a deficit of sources, with
3C\,390.3 as the closest possible source candidate. This deficit is not
a result of our selection criteria but rather a general property of the
distribution of radio galaxies, that shows a lack of sources at Supergalactic
longitudes between 0\degree and 90\degree~\cite{Peebles2021}.
Despite this deficit it is known since several decades~\cite{Shaver1990}
that radio galaxies tend to be close to the Supergalactic plane---such as
the hints for excesses in the UHECR data---whereas ordinary galaxies are
more isotropically distributed.
Finally, we note that the rather strong EGMF expected in the Supergalactic
plane allows for a scenario where some of the actual source of UHECRs
are no longer visible in the radio or even electromagnetic spectrum because of their finite lifetime but the majority of its delayed CRs are still arriving. 
This provides a possible but speculative scenario for the excess at about
($9^{\rm h}\,30^{\rm m}$, +54\degree) in particular.

\section{Conclusions}
\label{sec:concl}

We have examined if the stringent constraints imposed by the dipole and
quadropole anisotropies as well as the UHECR spectrum and composition
can be satisified by radio galaxies as UHECR sources. We have modeled
37~individual radio galaxies, constraining their
properties using information from the radio-CR correlation and a dynamical
evolution model. In addition, we have included the diffuse flux emitted by
the bulk of non-local radio galaxies based on their radio luminosity
distribution. Our approach of using a rigidity dependent injection
rate reduced the number of independent parameters considerably
and allowed us to fit the parameters $g_{\rm m}$ and $g_{\rm acc}$
individually for each source. Moreover, the CR propagation accounts
for deflections by a turbulent EGMF as well as the GMF model by JF12.

The observed dipole anisotropy imposes a  non-trivial constraint
on the distribution of local UHECR sources. In the case of radio
galaxies,  there is a significant over-fluctuation of these bright
sources at high Galactic longitudes ($l\gtrsim 240\degree$),
which is roughly aligned with the observed UHECR dipole.  Radio galaxies
could have been ruled out as the major UHECR source had this distribution been flipped by $\sim 180\degree$ in longitude.
In general, we have obtained a good description of the UHECR spectrum, its
composition and the dipole and quadrupole anisotropies, even using only
a small sample of the most powerful local sources combined with an isotropic
contribution from low-luminosity radio galaxies. In
particular, we find that scenarios where a few local sources---among them for instance Fornax~A, Virgo~A, 3C\,270, Centaurus~B or Hydra~A---dominate the flux above the ankle, while low-luminosity radio galaxies
contribute an isotropic background dominating below the ankle,
provide a good fit to the data. Moreover, we found that neither the closest
(Centaurus~A) nor the most powerful source (Cygnus~A) in the radio band
is likely to give a dominant contribution to the observed UHECR events.
Even though our results favor a major contribution by Fornax~A as well as 
Virgo~A and/or 3C\,270 additional data on medium- and large-scale anisotropies will be
necessary in order to draw clear conclusions on the importance of
individual local radio sources.

\acknowledgments

We would like to thank M.~Unger for fruitful discussions. Further, we are grateful to J. Machalski for applying his recent dynamical evolution model~\cite{Machalski+2021} on those
local, high-luminosity radio galaxies that became relevant in our work. BE acknowledges supported by the DFG grant EI~963/2-1.

\appendix
\section{Details on the optimization procedure}
\label{app:optProc}
One of the most critical issues that we had to tackle in our work is
the size of the parameter space and its adjustment during the fit procedure.
To ensure a fast performance the individual dipole vectors
$\vec{d}_s(E_1,\dots,E_4)\equiv d_{s,i}$ that result from the local sources
$s$ for a given angular distribution at Earth---as determined by the
concentration parameter $\kappa$---are calculated before starting the
optimization algorithm. Further, the optimization routine features a (fast)
inner linear least-squares problem solver that determines $g_{{\rm m},s}$ and
a (slower) outer global optimizer to determine $\alpha$,
$\tilde{g}_{{\rm acc},s}\equiv g_{{\rm acc},s}/\sqrt{1-g_{{\rm m},s}}$,
$g_{\rm m}^{\rm (l)/(h)}$ and $g_{\rm acc}^{\rm (l)/(h)}$.  This splitting can be
done since $\alpha$, $g_{\rm m}$ and $g_{\rm acc}$ only change the CR flux from
an individual, local source but not the angular distribution at Earth.
Therefore, only the norm of the pre-calculated vectors needs to be modified
by the optimization algorithm\footnote{We assume that $\alpha$ is the same for all sources but $\tilde{g}_{\rm acc}$ is allowed to differ.}. The impact
of the isotropic contribution from the bulk of radio sources is incorporated
by reducing the necessary contribution of local sources to the total,
isotropic flux by using $g_{\rm m}^{\rm (l)/(h)}$ and $g_{\rm acc}^{\rm (l)/(h)}$,
where we differentiate between low-luminosity (l) and high-luminosity (h)
sources. Thus, in total the algorithm optimizes $1+2\times 2+N_{\rm src}$
parameters (for $\alpha$, $g_{\rm m}^{\rm (l)/(h)}$, $g_{\rm acc}^{\rm (l)/(h)}$
and $\tilde{g}_{\rm acc}$), where $N_{\rm src}$ denotes the total number of
local sources, so that the chi-squared value (from the comparison with the
CR flux, the dipole strength and direction at different energies) becomes
minimal.

We first determine the non-normalized flux $\tilde{J}_s=J_s/g_{{\rm m},s}$ for
some values of $\alpha$ and
$\tilde{g}_{\rm acc}\equiv g_{\rm acc}/\sqrt{1-g_{\rm m}}$ and solve then
the linear least-squares problem 
\be
\vec{d}_{\rm obs}(E_1,\dots,E_4) \equiv d_{{\rm obs},i} = \sum_s \frac{J_s}{J_E}\, d_{s,i} \simeq \frac{\tilde{J}_s}{J_{\rm obs}}\, d_{s,i} \,g_{{\rm m},s}
\label{eq:linLSQ}
\ee
with bounds on the variables (using scipy\footnote{\url{https://docs.scipy.org/doc/scipy/reference/generated/scipy.optimize.lsq_linear.html}}) to obtain
the proper normalization $g_{\rm m}$ of the local sources based on the
observed dipole $\vec{d}_{\rm obs}(E_1,\dots,E_4)$ at four different energies.
Note that on the right hand side of Eq.~(\ref{eq:linLSQ}) we suppose that
the total isotropic flux $J_E=\sum_s J_s + J_{\rm csf}$ equals the observed
one $J_{\rm obs}$. This is only the case if the model provides also a good
fit to the observed, isotropic flux. However, this is a necessary condition
for a sufficiently small chi-squared value, since the flux uncertainty is
significantly smaller than the dipole uncertainty. 
This procedure yields the benefit of providing the $g_{{\rm m},s}$ values of
the local sources with less computational effort (especially in the case of
a large local source sample) as if one would include these parameters in the
optimization algorithm. 
Finally, we use $g_{{\rm m},s}$ as well as the other parameters to calculate
the total flux $J_E$ as well as the total dipole vectors $d_i$ and
determine the corresponding chi-squared value with respect to the
observational data. 

For the optimization routine by itself we tested different types of global
optimizers that are provided by the scipy package showing that the
differential evolution algorithm\footnote{\url{https://docs.scipy.org/doc/scipy/reference/generated/scipy.optimize.differential_evolution.html}} yields
the most robust results in our case.

\paragraph{Software:} Some of the results in this paper have been derived using the software packages Numpy \cite{vanDerWalt2011}, Scipy \cite{2020SciPy-NMeth}, Pandas \cite{mckinney-proc-scipy-2010}, Matplotlib \cite{Hunter:2007}, Seaborn \cite{michael_waskom_2017_883859} and HEALPix/ healpy \cite{2005ApJ...622..759G} as well as the CR simulation tool CRPropa\,3~\cite{CRPropa3_2016, CRPropa3.1_2017}.

\bibliographystyle{JHEP} 
\addcontentsline{toc}{section}{Bibliography}
\bibliography{references,ref2}

\end{document}